\documentclass[12pt]{article}

\usepackage{color}
\usepackage{epsfig,amssymb,amsfonts,amsmath,graphicx,dsfont,xfrac}
\usepackage{authblk}
\usepackage{subcaption}
\usepackage{epstopdf}
\definecolor{mygray}{gray}{0.5}

\usepackage{cite}
\usepackage[colorlinks=true,linkcolor=blue,citecolor=red]{hyperref}

\newcommand{\be}{\begin{equation}}
\newcommand{\ee}{\end{equation}}
\newcommand{\bea}{\begin{eqnarray}}
\newcommand{\eea}{\end{eqnarray}}

\title{Exact solutions for vector phase matching conditions in nonlinear uniaxial crystals}

\author[${}$]{Juan L\'opez-Dur\'an}
\author[${}$]{Oscar Rosas-Ortiz}

\affil[${}$]{\footnotesize Physics Department, Cinvestav, AP 14-740, 07000
M\'exico City, Mexico}

\date{}
\begin{document}

\maketitle

\begin{abstract}
The strongly transcendental equations of vector phase matching are transformed into a fourth order polynomial equation that admits analytical solution. The real roots of this equation provide the optical axis orientations that are useful for efficient down-conversion in nonlinear uniaxial crystals. The production of entangled photon pairs is discussed in both collinear and non-collinear configurations of the spontaneous parametric down conversion (SPDC) process. Degenerate and non-degenerate cases are also distinguished. As a practical example, SPDC processes of type-I and type-II are studied for beta-barium-borate (BBO) crystals. The predictions are in very good agreement with experimental measurements already reported in the literature, and include theoretical results of other authors as particular cases. Some properties that seem to be exclusive to BBO crystals are reported, the experimental verification of the latter would allow a better characterization of these crystals.
\end{abstract}


\section{Introduction}
\label{intro}
 
Entangled photon sources are critical for developing photonic quantum technologies \cite{Obr09} and for photonic quantum information processing \cite{Bou00}. These sources have also allowed for the experimental study of what the founding fathers of quantum theory liked to call thought experiments \cite{Har06,Fri18}. As a concept, entanglement appears in physics after intense debate. The term (loosely translated from the German word {\em verschr\"ankung}) was introduced by Schr\"odinger \cite{Sch35} to describe what occurs with our knowledge of two systems that are separated after they were interacting for a while, and from which we had maximal knowledge before the interaction \cite{Enr14}. As a result, instead of two isolated systems there is just a single composite system and therefore any change to one subsystem would affect the other, no matter the distance between them. The results obtained from interaction-free measurements \cite{Eli93,Kwi95a,Kwi96} are an indication that entanglement is indeed a fundamental property of the quantum systems. 

Bipartite entangled states can be prepared by producing an interaction between two different quantum systems in such a way that neither of the two emerging states has a definite value, but as soon as one of them is measured, the other state is automatically determined. In this context, the spontaneous parametric down-conversion (SPDC) process is a suitable way to produce entangled photon pairs. This occurs in nonlinear crystals, where one photon (pump) gives rise to a pair of entangled photons (signal and idler) \cite{Wal10,Cou18,Zha21}. SPDC is degenerate if the wavelength of the signal and idler photons are equal; otherwise, it is non-degenerate. Depending on the direction of propagation of the pump, signal and idler wave vectors, the SPDC process can be classified as collinear or non-collinear. The polarization of the new pair of photons characterize the SPDC process as follows. In type-I SPDC the polarization of the created photons is parallel to each other and orthogonal to the polarization of the pump photon. The light created in these conditions forms a cone aligned with the pump beam. In type-II SPDC the idler polarization is orthogonal to the signal one and the new light forms two cones that are not necessarily collinear. 

The type-II SPDC is particularly interesting as the photons produced are entangled in their polarization states \cite{Rub94,Rub96}, making them useful for representing qubits in quantum information \cite{Kwi95b}. One of the cones is ordinarily polarized and the other extraordinarily. Since they intersect in most configurations, it turns out that the generated light that propagates along these intersections is not polarized since we cannot distinguish if a certain photon belongs to one or another cone (possible entanglement is anticipated). Nevertheless, ordinary and extraordinary photons propagate with different speed inside the crystal \cite{Dmi99,Sal07}, so one photon comes out of the crystal before the other one \cite{Cou18}. Then, in principle, a time measurement can distinguish between the photon pairs along the intersections (meaning no entanglement) \cite{Bou00}. One gets quantum indistinguishability (entanglement) once the relative time ordering is compensated by using concrete arrays of crystals \cite{Deh02}. In this way, the entire Bell basis of bipartite entanglement can be achieved in terms of the polarization state of the photon pairs produced by type-II SPDC.

In general, SPDC follows conservation of energy and conservation of momentum, which are crucial for the process to occur \cite{Wal10,Cou18,Zha21}; the corresponding equations are called phase-matching conditions. In particular, conservation of the wave vector is required for an efficient non-linear effect, although it is an impossible condition to be satisfied with most materials \cite{Cou18}. The above condition is commonly achieved in birefringent nonlinear uniaxial crystals since they possess two different refractive indices along different symmetry axes for a given wavelength (biaxial crystals --with three different refractive indices-- are also available). 

To solve the phase-matching conditions in vector form, it is convenient to work in spherical coordinates. Then, nine parameters are to be determined: three wavelengths (conservation of energy) and three pairs of polar and azimuthal angles (conservation of momentum). However, for light propagating within nonlinear crystals, depending on whether it is polarized ordinary or extraordinarily, the refractive index is expressed in terms of both the wavelength and the angle formed by the wave vector and the corresponding optical axis \cite{Dmi99,Sal07}. This makes obtaining analytical solutions for phase matching conditions a surprisingly difficult task, especially in the non-collinear case if one is looking for the production of entangled photon pairs. 

Since the refractive index of extraordinarily polarized light is a very elaborated function of the unknowns, and the latter are encapsulated by trigonometric functions, determining the parameters requires solving strongly transcendental equations. A fact that has motivated more the study of numerical approximations than the search for analytical solutions \cite{Bat73,Boe00,Cou18}.

In this work we show that the difficulty of solving the strongly transcendental equations of vector phase-matching is reduced by transforming them into a fourth-order polynomial equation that admits analytical solution. The corresponding roots are complex-valued in general, so we impose a reality condition that determines the optical axis orientations that are useful for efficient down-conversion. Our research is addressed to the type-II SPDC process in nonlinear uniaxial crystals, including type-I SPDC as particular case, with emphasis in the non-collinear case. 

The usefulness of the analytical solutions reported here is twofold: they contribute to a better understanding of the SPDC process by expanding the set of exactly solvable cases, and are helpful in the design of entangled photon sources. 

To provide a practical example we consider the nonlinear crystal beta-barium-borate (BBO), which is negative uniaxial (although the approach can include the properties of biaxial crystals). Our results are in complete agreement with theoretical and experimental work already reported by other authors, and include some refinements whose full experimental verification remains an open question.

The remainder of the paper is organized as follows. In Section~\ref{sec2}, we introduce some basic notions of the phase-matching conditions and establish the problem to be solved. The vectorial conditions for phase-matching are simplified to a system of two coupled equations for the polar angles of idler and signal photons. For type-II SPDC, these equations are transformed into a fourth-order polynomial equation whose solutions provide the polar angle of the idler beam. In Section~\ref{sec3} we provide the exact solution for such equation and particularize to the SPDC process in a BBO crystal. We analyze both degenerate and non-degenerate cases. Five general configurations of the cones of down-converted light are discussed, they include beam-like, divergent, osculating (entanglement in collinear beams), overlapping (entanglement in spatially separate beams), and coaxial cones. In Section~\ref{sec4}, we discuss our results by comparing them with the work of other authors. Our theoretical model is in close agreement with experimental measurements and theoretical approaches already reported in the literature. Finally, Appendix~\ref{ApA} includes some concrete calculations that are useful to follow the discussion throughout the manuscript.

\section{Laws of conservation for SPDC}
\label{sec2}

The frequency-matching and phase-matching conditions of spontaneous parametric down conversion (SPDC) are respectively written as
\be
\omega_p =\omega_s + \omega_i, \qquad {\mathbf k}_p= {\mathbf k}_s + {\mathbf k}_i,
\label{pm}
\ee
where $\omega_u$ and ${\mathbf k}_u$ are the angular frequency and wave-vector of the $u$-light wave, with $u=p,s,i$, referring to pump, signal and idler fields. 

The conditions (\ref{pm}) arise from the temporal and spatial phase matching of the waves associated with the three fields in SPDC phenomena, and ensure the mutual interaction of the fields over extended durations of time and regions of space \cite{Sal07}. In general, they lead to multiple solutions where the down-converted light takes the form of a cone of multispectral light \cite{Sal07}.

We are interested in finding exact solutions to the phase-matching condition for both types of nonlinear uniaxial crystals, I and II, and for two general cases (degenerate and non-degenerate) of the frequency-matching condition. That is, our program will hold for any relationship between the angular frequencies, whenever they satisfy the frequency-matching (\ref{pm}). In this form, we shall assume that the values of $\omega_p$, $\omega_s$, and $\omega_i$ are available from either direct measurements in the laboratory or appropriate theoretical considerations. 

The reference system in laboratory is defined by the cartesian unitary vectors in $\mathbb R^3$: $\hat {\mathbf e}_1 \equiv \hat {\mathbf e}_x$, $\hat {\mathbf e}_2 \equiv \hat {\mathbf e}_y$, and $\hat {\mathbf e}_3 \equiv \hat {\mathbf e}_z$. The coordinates are right-handed, with axes $x_1 \equiv x$, $x_2 \equiv y$ and $x_3 =z$. The origin of coordinates is located at the center of mass of the crystal, which will be considered a rectangular cuboid for simplicity.

\begin{figure}
\centering
\includegraphics[width=0.3\textwidth]{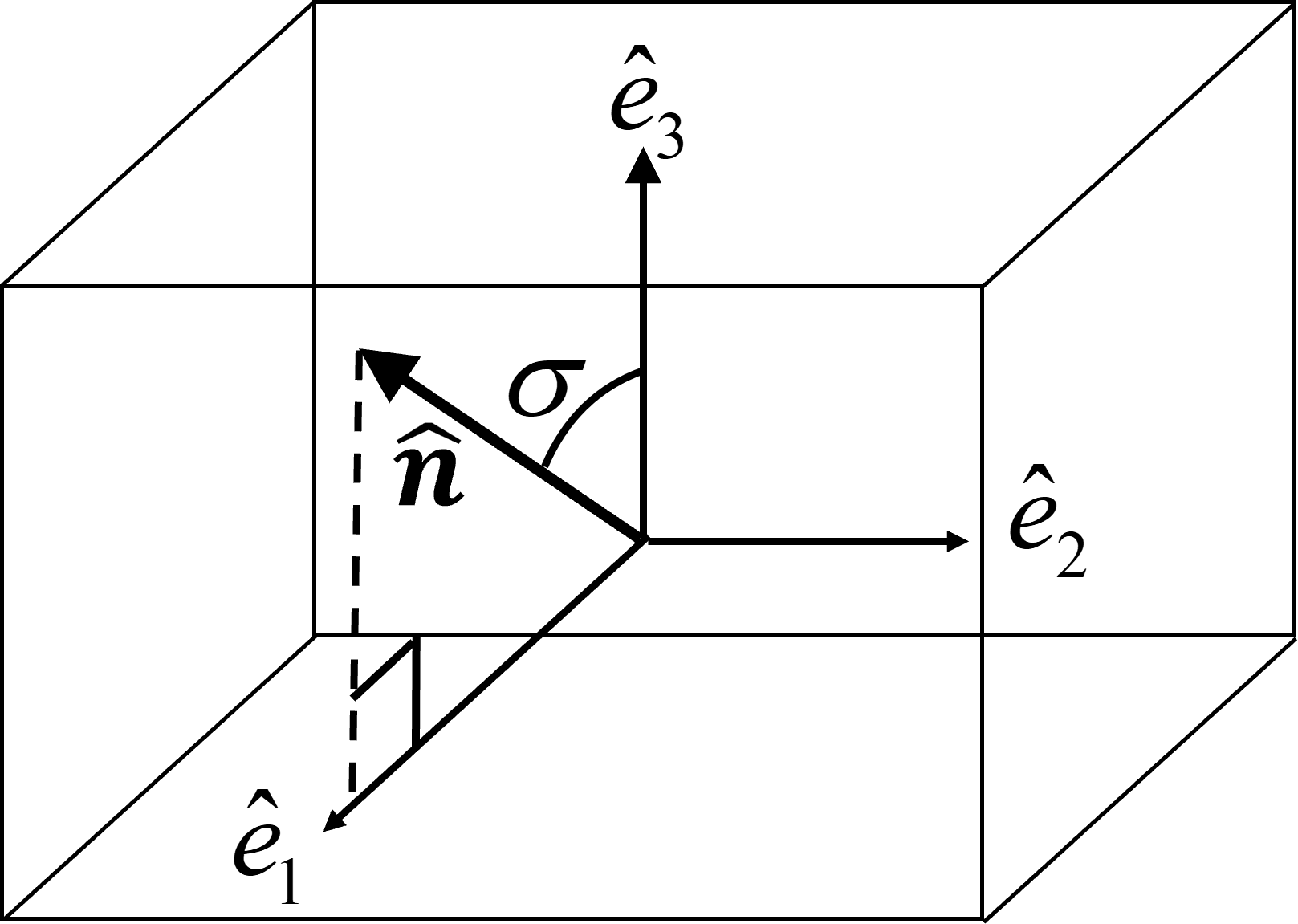}
\caption{\footnotesize The cartesian unitary vectors $\hat {\mathbf e}_k \in \mathbb R^3$, with $k=1,2,3$, define the reference system in laboratory, where the optic axis of the nonlinear crystal lies in the $xz$-plane and forms the angle $\sigma$ with the $z$-axis. Thus, the optic axis is characterized by the unitary vector $\hat {\mathbf n}=  \hat {\mathbf e}_1  \sin \sigma +  \hat {\mathbf e}_3  \cos \sigma$, with $\sigma \in [0, \pi]$. 
}
\label{fig01}
\end{figure}

Without loss of generality, we will assume that the optic axis of the crystal is oriented according to the unitary vector $\hat {\mathbf n}=  \hat {\mathbf e}_1  \sin \sigma +  \hat {\mathbf e}_3  \cos \sigma$, with $\sigma \in [0, \pi]$, see Figure~\ref{fig01}. Additionally, we shall consider that the pump-light wave propagates along the $z$-axis.

In general, to satisfy the phase-matching condition (\ref{pm}), the wave-vectors $\mathbf k_u$ obey the parallelogram rule of vector addition. Then we talk about {\em vector phase-matching}. However, depending on the applications of the down-converted light, one might be interested in studying only those vectors $\mathbf k_u$ that satisfy the additional condition of being collinear $\mathbf k_u = k_u \hat{\mathbf e}_0$, with $\hat{\mathbf e}_0$ a unitary vector in $\mathbb R^3$. In such a case we talk about {\em scalar} (or {\em collinear}) {\em phase-matching}. Our approach faces the problem in general (vector) form. Once this is solved exactly, we particularize to the simplest (scalar) form by adjusting the parameters of the general solution. 

The down-converted light is expected to form cones whose vertices lie inside the crystal. For the sake of simplicity we shall assume that all vertices coincide with the origin of coordinates in the laboratory frame. 

\begin{figure}
\centering
\includegraphics[width=0.4\textwidth]{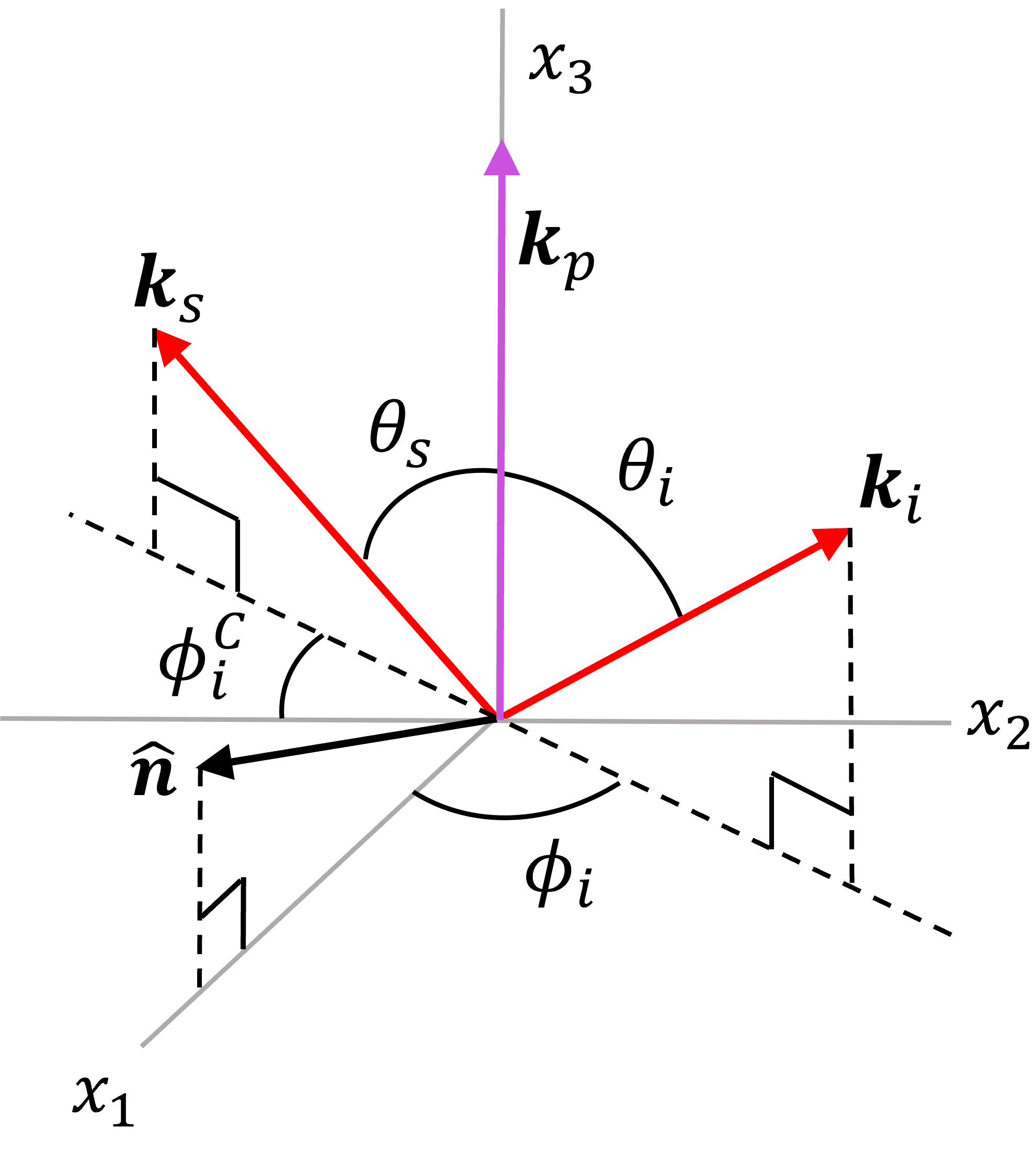}
\caption{\footnotesize The wave-vectors in spherical coordinates. The configuration assumes conservation of momentum ${\mathbf k}_p= {\mathbf k}_s + {\mathbf k}_i$. Here $x_k$, with  $k=1,2,3$, stand for cartesian axes in $\mathbb R^3$, and $\phi_i^C$ is complementary to $\phi_i$ in the third quadrant of the $xy$-plane. 
}
\label{fig02}
\end{figure}

Therefore, using spherical coordinates, the wave-vectors are written as follows
\be
\begin{array}{l}
{\mathbf k}_p = k_p (0,0,1), \\[1ex]
{\mathbf k}_s= k_s (\sin \theta_s \cos \phi_s, \sin \theta_s \sin \phi_s, \cos \theta_s), \\[1ex]
{\mathbf k}_i= k_i (\sin \theta_i \cos \phi_i, \sin \theta_i \sin \phi_i, \cos \theta_i),
\end{array}
\label{kas}
\ee
where $\theta_u \in [0, \pi]$ and $\phi_u \in [0, 2 \pi)$ stand for the polar and azimuthal angles, respectively, and $k_u = \Vert {\mathbf k}_u  \Vert$ refers to the wave-number of $u$-light wave, see Figure~\ref{fig02}.

The problem is to determine the wave-numbers $k_s$, $k_i$, as well as the angles $\theta_s$, $\phi_s$, and $\theta_i$, $\phi_i$, such that the phase-matching (\ref{pm}) is satisfied by giving $k_p$, $\omega_p$, and $\sigma$ (remember that we are assuming frequency-matching is assured).

Introducing (\ref{kas}) into the phase-matching (\ref{pm}) leads to the system
\be
k_s \sin \theta_s = - k_i \sin \theta_i \frac{\cos \phi_i}{\cos \phi_s} = - k_i \sin \theta_i \frac{\sin \phi_i}{\sin \phi_s},
\label{pm1}
\ee
\be
k_p = k_s \cos \theta_s + k_i \cos \theta_i.
\label{pm2}
\ee
From (\ref{pm1}) one immediately obtains $\tan \phi_s = \tan \phi_i$, which means a $\pi$-shift between the azimuthal angles
\be
\phi_s = \phi_i + \pi
\label{pm3}
\ee
that defines the relationship between the polar angles as follows
\be
k_s \sin \theta_s =  k_i \sin \theta_i.
\label{pm4}
\ee
The roots of the system formed by Eqs.~(\ref{pm2})--(\ref{pm4}) ensure the conservation of linear momentum in the SPDC process. 

On the other hand, for uniaxial crystals with ordinary and extraordinary refractive indexes, $n_O(\omega)$ and $n_E(\omega)$, to describe the propagation of ordinary light we require only the frequency dependent refractive index $n(\omega) =n_O(\omega)$. However, for extraordinary light waves, the refractive index depends also on the angle $\delta$ formed by the wave-vector and the optic axis according to the expression \cite{Sal07}
\be
\frac{1}{n^2(\omega; \delta)} = \frac{\cos^2 \delta}{n_O^2(\omega)} + \frac{\sin^2 \delta}{n_E^2(\omega)}.
\label{index}
\ee
The indexes $n_O(\omega)$ and $n_E(\omega)$ are determined by the Sellmeier equations \cite{Dmi99,Sal07}.

Remark that Eq.~(\ref{index}) introduces additional degrees of freedom to the problem we are dealing with. Indeed, each of the three waves of light can be ordinary or extraordinary. Then, according to the polarization of the $u$-wave, the refractive index $n(\omega_u; \delta_u)$ depends on the angle $\delta_u$  that is formed by the wave-vector ${\mathbf k}_u$ and the optic axis. Using the inner product of ${\mathbf k}_u$ with the unitary vector $\hat {\mathbf n}$ that characterizes the optic axis one has
\[
\cos \delta_u= \frac{\mathbf k_u \cdot \hat {\mathbf n}}{k_u}, \quad u= p,s,i.
\]
Then $\delta_p =\sigma$, and
\be
\cos \delta_u=  \sin \theta_u \cos \phi_u \sin \sigma + \cos \theta_u \cos \sigma, \quad u= s,i.
\label{del1}
\ee
That is, the angle formed by the idler and signal beams with the optic axis depends on $\sigma$ as well as on the phase-matching angles we are looking for.

Therefore, according to the polarization of idler and signal waves, the refractive index for these waves could depend on $\sigma$ and the corresponding phase-matching angles. If the polarizations of both light waves are equal (orthogonal) then the SPDC phenomenon is of type~I~(II) \cite{Sal07}. We are going to solve exactly the phase-matching for both types.

The number of unknowns in equations (\ref{pm2}) and (\ref{pm4}) can be reduced by expressing the wave-numbers $k_u$ in terms of the frequency and refractive index
\be
k_u= \frac{\omega_u}{c_0} n(\omega_u; \delta_u), \quad u=p,s,i,
\label{wn}
\ee
where $c_0$ is the speed of light in vacuum. Therefore
\be
\omega_p n_p (\omega_p; \sigma) = \omega_s  n_s(\omega_s; \delta_s) \cos \theta_s + \omega_i  n_i(\omega_i; \delta_i) \cos \theta_i,
\label{g3}
\ee
\be
\omega_s n_s(\omega_s; \delta_s) \sin \theta_s =  \omega_i n_i(\omega_i; \delta_i) \sin \theta_i,
\label{g1}
\ee 
so we just need to determine the angles $\theta_s$ and $\theta_i$ (remember that $\phi_s$ is the $\pi$-shifted version of $\phi_i$). That is, we have simplified the vector phase-matching (\ref{pm}) to the system of scalar equations composed of (\ref{pm3}), (\ref{g3}) and (\ref{g1}).  They, together with the frequency-matching condition (\ref{pm}), must be simultaneously satisfied. 

However, by reducing the number of unknowns we increase the complexity of the problem because, according to Eq.~(\ref{index}), the relationship between the refractive index $ n(\omega_u; \delta_u)$ and the angle $\delta_u$ is not only quadratic but transcendental for extraordinary $u$-waves. In turn, Eq.~(\ref{del1}) connects $\delta_u$ with the unknowns $\theta_u$ and $\phi_u$ in transcendental form, no matter the polarization of the $u$-wave. Then, depending on the polarization of the down-converted waves, the system (\ref{g3})-(\ref{g1}) could include transcendental equations of at least second degree.

Clearly, solving the pair (\ref{g3})-(\ref{g1}) requires concrete information about the character of the fields as they propagate in the crystal. For clarity, we shall analyze the phase-matching of types~I and II separately.

\subsection{Type-I SPDC}

For nonlinear uniaxial crystals of type~I, the idler and signal fields are polarized in ordinary form. We write $n_O(\omega_i)$ and $n_O(\omega_s)$ for the respective refractive indexes. In turn, the polarization of the pump-field is extraordinary, so the refractive index is a function of $\delta_p =\sigma$ (the angle between $\mathbf k_p$ and $\hat {\mathbf n}$), written from (\ref{index}) as follows
\be
n_p (\omega_p; \sigma) = \frac{n_O(\omega_p)}{\sqrt{1+ \left( \frac{n_O^2(\omega_p)}{n_E^2(\omega_p)} -1 \right) \sin^2 \sigma}}.
\label{pump1}
\ee
That is, when the pump beam enters the crystal forming the angle $\sigma$ with the optical axis, the extraordinary index will be different for different values of $\sigma$.

The function $n_p (\omega_p; \sigma)$ is positive for any value of $\sigma \in [0, \pi]$, and satisfies
\[
\left. n_p (\omega_p; \sigma) \right\vert_{\sigma=0^\circ} = \left. n_p (\omega_p; \sigma) \right\vert_{\sigma= 180^\circ} =n_O(\omega_p), \qquad \left. n_p (\omega_p; \sigma) \right\vert_{\sigma= 90^\circ} = n_E(\omega_p).
\] 
With (\ref{pump1}), equations~(\ref{g3}) and (\ref{g1}) acquire the form
\be
\omega_p n_p (\omega_p; \sigma) = \omega_i  n_O(\omega_i) \cos \theta_i + \omega_s  n_O(\omega_s) \cos \theta_s, \quad 
\omega_s n_O(\omega_s) \sin \theta_s = \omega_i n_O(\omega_i) \sin \theta_i.
\label{t1g}
\ee
Squaring both equations of (\ref{t1g}), after some simplifications, we obtain 
\be
\theta_i = \arccos \left[
\frac{\omega_p^2 n_p^2 (\omega_p; \sigma) + \omega_i^2 n_O^2(\omega_i) -\omega_s^2 n_O^2(\omega_s) 
}{2\omega_p \omega_i n_p (\omega_p; \sigma) n_O (\omega_i)}
\right].
\label{known}
\ee
The signal angle $\theta_s$ is derivable from (\ref{known}) and the second equation of (\ref{t1g}).

In the degenerate configuration ($\omega_s=\omega_i=\omega_p/2$), the above results are simplified as follows
\[
\theta_s = \theta_i =\arccos \left[ \frac{n_p (\omega_p; \sigma)}{n_O (\omega_p/2)} \right].
\]
Thus, the light produced by (degenerate) type-I SPDC describes a right circular cone, the axis of which is along the propagation direction of the pump beam, with aperture $2\theta_i = 2\theta_s$ (inside the crystal). In addition, the $\pi$-shift between azimuthal angles (\ref{pm3}) means that the created photon pairs are emitted on opposite sides of the corresponding cone. 

\subsection{Type-II SPDC}

In the case of nonlinear uniaxial crystals of type~II, consider the situation in which the signal-field is ordinarily polarized. The pump-field is still associated with the refractive index (\ref{pump1}), and the idler-field is now linked to the function
\be
n_i (\omega_i; \delta_i) = \frac{n_O(\omega_i)}{\sqrt{1+ \left( \frac{n_O^2(\omega_i)}{n_E^2(\omega_i)} -1 \right) \sin^2 \delta_i}},
\label{pump2}
\ee
which is positive for any value of $\delta_i \in [0, \pi]$ fulfilling (\ref{del1}), and satisfies
\[
\left. n_i (\omega_i; \delta_i) \right\vert_{\delta_i = 0^\circ}= \left. n_i (\omega_i; \delta_i) \right\vert_{\delta_i = 180^\circ}= n_O(\omega_i), \qquad \left. n_i (\omega_i; \delta_i) \right\vert_{\delta_i = 90^\circ} = n_E(\omega_i).
\] 
Using (\ref{pump2}), equations (\ref{g3}) and (\ref{g1}) are rewritten as follows:
\be
\omega_p n_p (\omega_p; \sigma) = \omega_s  n_O(\omega_s) \cos \theta_s + \omega_i  n_i(\omega_i; \delta_i) \cos \theta_i,
\label{rev1}
\ee
\be
\omega_s n_O(\omega_s) \sin \theta_s =  \omega_i n_i(\omega_i; \delta_i) \sin \theta_i.
\label{ref0}
\ee
After squaring (\ref{rev1}) and (\ref{ref0}), we may solve the system by preserving $\theta_i$. The straightforward calculation gives rise to the quadratic form 
\be
\omega_i^2 n_i^2 (\omega_i; \delta_i) - 2 \omega_i \omega_p n_i(\omega_i; \delta_i) n_p(\omega_p; \sigma) \cos \theta_i 
- \omega_s^2 n_O^2 (\omega_s) + \omega_p^2 n_p^2(\omega_p; \sigma) =0.
\label{rev1b}
\ee
Our program is completed after solving the transcendental equation~(\ref{rev1b}) for the phase-matching angle $\theta_i$. 

In the previous sections we have emphasized that solving equations like (\ref{rev1b}) is much more difficult than it seems at first glance. Indeed, according to (\ref{pump2}), the refractive index $n_i(\omega_i; \delta_i)$ is a very elaborated function of $\theta_i$, so  (\ref{rev1b}) is strongly transcendental. This fact could motivate more the study of numerical approaches than the search for analytical solutions \cite{Bat73,Boe00,Cou18}. However, we are going to show that the complexity of solving (\ref{rev1b}) is reduced by transforming it into a fourth-order polynomial equation.

\subsubsection{Fourth-order polynomial equation for type-II SPDC}

To simplify calculations, it is useful to square (\ref{del1}) in the form
\[
\cos^{2}\delta_i = \dfrac{T}{1+\tan^{2}\theta_{i}}, 
\]
where
\[
T=\cos^{2}\phi_i \sin^{2}\sigma \tan^{2}\theta_{i}+2 \sin \sigma \cos \sigma \cos \phi_i  \tan\theta_{i} + \cos^{2}\sigma.
\]
Then, the refractive index (\ref{pump2}) can be expressed as follows
\be
\frac{1}{n_i^{2}\left(\omega_{i}; \delta_i \right)} = \frac{1 + \tan^2 \theta_i + n_E^2(\omega_i) T \Delta
}{(1 + \tan^2 \theta_i) n_E^2(\omega_i)},
\label{clave}
\ee
with
\be
\Delta = \frac{1}{n_O^{2} \left( \omega_{i} \right) } - \frac{1}{n_E^{2} \left( \omega_{i} \right)}.
\label{cdel}
\ee
In turn, (\ref{rev1b}) can be simplified in the form
\be
\xi^2  - 2\xi \frac{n_{p}\left(\omega_p; \sigma \right) }{n_i \left( \omega_{i}; \delta_i \right)} \cos \theta_{i} - \frac{ \gamma}{ n_i^2 \left(\omega_{i}; \delta_i \right)} =  0,
\label{rev2a}
\ee
where
\[
\xi = \frac{\omega_i}{\omega_p}, \qquad 
\gamma = \frac{\omega^2_s}{\omega_p^2} n^{2}_O \left(\omega_{s} \right) -  n^{2}_{p} \left( \omega_p; \sigma \right).
\]
To avoid the square-root appearing in $n_i(\omega_i; \delta_i)$, let us square (\ref{rev2a}) to arrive at the following quartic form for $\xi$,
\be
\xi^4 -2 \xi^2\left[ \frac{ \gamma + 2 n_p^2(\omega; \sigma) \cos^2 \theta_i}{n_i^2 (\omega_i; \delta_i)} \right] + \frac{\gamma^2}{n_i^4(\omega_i; \delta_i)}=0.
\label{quartic}
\ee
From (\ref{clave}), it is clear that (\ref{quartic}) is indeed a fourth-order polynomial equation in the variable 
\be
\chi = \tan \theta_i.
\label{new}
\ee 
The straightforward calculation yields
\be
a_4 \chi^4 + a_3 \chi^3 + a_2 \chi^2 + a_1 \chi +a_0 =0, 
\label{taneq}
\ee
where the coefficients $a_k$, $k=0,1,2,3,4$, are real-valued functions of the azimuthal angle $\phi_i$, the orientation $\sigma$ of the optic axis, and the three angular frequencies $\omega_p$, $\omega_s$, $\omega_i$, see Appendix~\ref{ApA} for details.

\section{Exact solution of the SPDC phase matching conditions for crystals of  type~II}
\label{sec3}

The complete solution to vector part of Eq.~(\ref{pm}) for nonlinear uniaxial crystals of type~II is obtained after solving (\ref{taneq}). Indeed, we determine the polar angle $\theta_i$ by reversing (\ref{new}) with $\chi$ the appropriate root of Eq.~(\ref{taneq}). The other polar angle $\theta_s$ is obtained from either (\ref{rev1}) or  (\ref{ref0}). In turn, as indicated above, $\phi_s$ is uniquely determined by $\phi_i$ through the $\pi$-shift (\ref{pm3}). 

The fourth-order polynomial equation (\ref{taneq}) is better studied in its monic form
\be
\chi^{4}+b_{3} \chi^{3}+b_{2} \chi^{2}+b_{1}\chi+b_{0} = 0,
\label{monic}
\ee
with $b_k = a_k/a_4$, $k=0,1,2,3$. In this terms, the four roots are as follows 
\be
\begin{array}{c}
\chi_1 = - \frac12 \sqrt{\eta} + \Gamma_+ - \frac14 b_3, \qquad \chi_2 = -\frac12 \sqrt{\eta} - \Gamma_+ - \frac14 b_3,\\[2ex]
\chi_3 = \frac12 \sqrt{\eta} + \Gamma_- - \frac14 b_3, \qquad \chi_4 =\frac12 \sqrt{\eta} - \Gamma_- - \frac14 b_3,
\end{array}
\label{set}
\ee
where 
\[
\Gamma_{\pm} = \frac12 \left[ -2  p - \eta  \pm  2 q \eta^{-1/2}  \right]^{1/2}, \quad
p =b_{2} -\tfrac38 b_3^2, \quad q =\tfrac18 b_3^3 -\tfrac12 b_{3} b_{2} +b_{1},
\]
and $\eta$ is a root of the cubic equation
\be
\eta^{3} +2p \eta^{2} + ( p^{2} -4r ) \eta - q^{2} =0,
\label{para3}
\ee
with
\[
r = - \tfrac{3}{256}b_{3}^{4} + \tfrac{1}{16} b_{3}^{2} b_{2} -\tfrac{1}{4}b_{3} b_{1} +b_{0}.
\]
For a detailed derivation of the previous formulae see Appendix~\ref{ApA}.

$\bullet$ {\bf Refraction.} The results obtained above (and in the previous sections) refer to light propagating within the crystal. A more realistic model should consider that the detection zone is far from the crystal, where the light to be detected has undergone refraction. That is, we have to take into account the transition from the crystal to the medium in which the detectors --and the crystal itself-- are embedded. For simplicity, we will suppose the down-converted light to propagate in air ($n=1$) as soon as it leaves the crystal. 

Assuming that the normal to the interface is parallel to $\hat{\mathbf e}_3$, see Figure~\ref{fig01}, for type-II SPDC the Snell law yields
\[
n_O( \omega_s ) \sin \theta_s = \sin \theta_s^{(a)}, \qquad n_i (\omega_i; \delta_i ) \sin \theta_i = \sin \theta_i^{(a)},  
\]
where $\theta_s^{(a)}$ and $\theta_i^{(a)}$ are the polar angles of refraction for signal and idler waves, respectively. From Eq.~(\ref{ref0}), one immediately obtains 
\be
\sin \theta_s^{(a)} =  \frac{\omega_i}{\omega_s} \sin \theta_i^{(a)}.
\label{snell2}
\ee
Therefore, after refraction, the relationship between the polar angles of down-converted light  is uniquely determined by the way in which the frequency-matching ($\omega_p = \omega_s + \omega_i$) is satisfied.

In particular, for degenerate frequencies $\omega_i = \omega_s  \, (= \omega_p/2)$, one has $\theta_s^{(a)} = \theta_i^{(a)}$. That is, if the down-converted light is produced in degenerate form, signal and idler cones will be observed with the same inclination with respect to the pump beam, and will have the same aperture. Other combinations of $\omega_i$ and $\omega_s$ fulfilling the frequency-matching lead to different configurations of the cones at the detection zone, as we are going to see.

In the sequel, we will emphasize the predictions after refraction since they are the subject of interest for experimental data in laboratory. 

$\bullet$ {\bf Detection plane.} Without detection, the down-converted cones extend infinitely far. Their conical surfaces are formed by half-lines (generatrix lines) whose orientation is determined by the wave-vectors $\mathbf k_s$ and $\mathbf k_i$. 

Positioning a transversal detection plane $\Sigma_0$ at $z=z_0 > L$, with $L$ the crystal length, the directrix (base) of each cone is defined by the intersection with $\Sigma_0$. The lateral surfaces are then formed by line segments (generatrix lines) that are oriented according to $\mathbf k_s$ and $\mathbf k_i$.  Therefore, in laboratory, the most general configuration of each cone is oblique circular, where the axis is not orthogonal to the base (a circle), so the base center does not coincide with the projection of the apex on $\Sigma_0$. As we are going to see, right circular cones are also allowed, but they require a very concrete orientation $\sigma$ of the optic axis as well as particular configurations of the frequency-matching.

We assume that it is feasible to operate with detectors that can be placed (and moved) along the detection plane $\Sigma_0$, which is indeed a displaced version of the $xy$-plane of the laboratory frame shown in Figure~\ref{fig01}. The plane $\Sigma_0$ is  where the transverse patterns of both cones are to be formed.

$\bullet$ {\bf BBO crystal.} To provide a practical example we will consider the nonlinear crystal beta barium borate (BBO, $\beta\operatorname{-BaB}_2 \!\operatorname{O}_4$), which is negative uniaxial $(n_O > n_E)$ and is characterized by the wavelength-dependent refractive indexes \cite{Dmi99,Sal07}:
\bea
n_O^2 (\lambda) =2.7359 + \frac{0.01878}{\lambda^2 - 0.01822}  - 0.01354 \lambda^2,
\label{uno}\\[1ex]
n_E^2(\lambda) = 2.3753 + \frac{0.01224}{\lambda^2 - 0.01667} - 0.0516 \lambda^2.
\label{dos}
\eea
The Sellmeier equations (\ref{uno})-(\ref{dos}) consider refractive indexes at room temperature, with wavelengths $\lambda$ measured in $\mu m$, and  are valid in the wavelength range (0.22--1.06) $\mu$m \cite{Sal07}. The ordinary refractive index (\ref{uno}) is larger than the extraordinary one (\ref{dos}) at any allowed wavelength $\lambda$. 

Using a BBO crystal in the laboratory, the phase-matching of both types~I $(\operatorname{e} \rightarrow \operatorname{o} + \operatorname{o})$ and II $(\operatorname{e} \rightarrow \operatorname{e} + \operatorname{o}, \operatorname{e} \rightarrow \operatorname{o} + \operatorname{e})$ can be achieved by tuning the angle $\sigma$ formed by the optical axis and the pump beam, so this crystal is suitable for testing our theoretical results.

The BBO crystal belongs to the symmetry group 3m \cite{Mid65,Eim87,Nik91,Dmi99,Men07}.  Considering the second-order nonlinear coefficients $d_{i,J}$ for this group, the straightforward calculation yields three different expressions for the effective nonlinearity $d_{\operatorname{eff}}$ of BBO crystals. For type-I BBO, taking into account the anisotropy of the downconverter and the allowed polarizations of the medium, we obtain
\be
d_{\operatorname{eff}}^{\operatorname{ooe}} = \frac{- \varepsilon_{po} d_{31}  \cos (\varphi_i -\varphi_s) \sin \vartheta_p + \varepsilon_{pe} d_{22}  \sin (\varphi_i + \varphi_s + \varphi_p) \cos \vartheta_p}{\left( \varepsilon_{pe}^2 \cos^2 \vartheta_p + \varepsilon_{po}^2 \sin^2 \vartheta_p \right)^{1/2}},
\label{juan1}
\ee
where $\varepsilon_{ue}= n^2_{ue}$ and $\varepsilon_{uo}= n^2_{uo}$ stand for the ordinary and extraordinary relative dielectric constants of the medium, with $n_{ue}$ and $n_{uo}$ the extraordinary and ordinary refractive indexes for the $u$-wave. The three-index notation $d_{\operatorname{eff}}^{\square \square \square}$ designates the polarization of idler, signal, and pump waves in that order. At the limit $\varepsilon_o \rightarrow \varepsilon_e$, up to a global $(-)$ we have
\be
\lim_{\varepsilon_o \rightarrow \varepsilon_e} d_{\operatorname{eff}}^{\operatorname{ooe}} = d_{31}  \cos (\varphi_i -\varphi_s) \sin \vartheta_p - d_{22}  \sin (\varphi_i + \varphi_s + \varphi_p) \cos \vartheta_p.
\label{type1gral}
\ee
In turn, for type-II BBO we arrive at the formulae
\be
\begin{array}{rl}
d_{\operatorname{eff}}^{\operatorname{eoe}} = 
& \left[ \left( \varepsilon_{pe}^2 \cos^2 \vartheta_p + \varepsilon_{po}^2 \sin^2 \vartheta_p \right) \left( \varepsilon_{ie}^2 \cos^2 \vartheta_i + \varepsilon_{io}^2 \sin^2 \vartheta_i \right) \right]^{-1/2} \times \\[1ex]
& \quad \left\{ \varepsilon_{pe} \left[ \varepsilon_{ie} d_{22} \cos \vartheta_i \cos (\varphi_i + \varphi_s + \varphi_p) - \varepsilon_{io}d_{31}  \sin \vartheta_i \sin (\varphi_s - \varphi_p)  \right] \cos \vartheta_p  
\right.\\[1ex] 
& \quad \quad \left.  +  \, \varepsilon_{po} \varepsilon_{io}d_{31} \cos \vartheta_i \sin (\varphi_i - \varphi_s)  \sin \vartheta_p \right\},
\end{array}
\label{juan2}
\ee
and 
\be
\begin{array}{rl}
d_{\operatorname{eff}}^{\operatorname{oee}} = 
& \left[ \left( \varepsilon_{pe}^2 \cos^2 \vartheta_p + \varepsilon_{po}^2 \sin^2 \vartheta_p \right) \left( \varepsilon_{se}^2 \cos^2 \vartheta_i + \varepsilon_{so}^2 \sin^2 \vartheta_i \right) \right]^{-1/2} \times \\[1ex]
& \quad \left\{ \varepsilon_{pe} \left[ \varepsilon_{se} d_{22} \cos \vartheta_s \cos (\varphi_i + \varphi_s + \varphi_p) - \varepsilon_{so}d_{31}  \sin \vartheta_s \sin (\varphi_i - \varphi_p)  \right] \cos \vartheta_p  
\right.\\[1ex] 
& \quad \quad \left.  -  \, \varepsilon_{po} \varepsilon_{se}d_{31} \cos \vartheta_s \sin (\varphi_i - \varphi_s)  \sin \vartheta_p \right\}.
\end{array}
\label{juan3}
\ee
At the isotropic limit $\varepsilon_o \rightarrow \varepsilon_e$, the above expressions become
\be
\begin{array}{rl}
\lim_{\varepsilon_o \rightarrow \varepsilon_e} d_{\operatorname{eff}}^{\operatorname{eoe}} =  
& \left[  d_{22} \cos \vartheta_i \cos (\varphi_i + \varphi_s + \varphi_p) - d_{31}  \sin \vartheta_i \sin (\varphi_s - \varphi_p)  \right] \cos \vartheta_p  \\[1ex] 
&   +  d_{31} \cos \vartheta_i \sin (\varphi_i - \varphi_s)  \sin \vartheta_p,
\end{array}
\label{eoe}
\ee
and
\be
\begin{array}{rl}
\lim_{\varepsilon_o \rightarrow \varepsilon_e} d_{\operatorname{eff}}^{\operatorname{oee}} = 
& \left[ d_{22}  \cos \vartheta_s  \cos (\varphi_i + \varphi_s + \varphi_p)  - d_{31} \sin \vartheta_s \sin (\varphi_i - \varphi_p) \right] \cos \vartheta_p  \\[1ex]
& - d_{31} \cos \vartheta_s  \sin (\varphi_i - \varphi_s) \sin \vartheta_p.
\end{array}
\label{oee}
\ee
The polar and azimuthal angles $(\vartheta_u, \varphi_u)$ are measured with respect to the crystal reference frame, where the optical axis is along the corresponding $z$-axis.

The above expressions for the effective nonlinearity of BBO crystals refer to the vector phase-matching, where the wave-vectors $\mathbf k_u$ obey the parallelogram rule to satisfy the vector addition (\ref{pm}). To our knowledge, equations (\ref{juan1})-(\ref{oee}), have not been previously reported. 

In contrast, what is well known is the calculation of the effective nonlinearity for scalar (or collinear) phase-matching. In this case the three pairs of angles $(\vartheta_u, \varphi_u)$ satisfy $\vartheta_u = \sigma$ and $\varphi_u = \varphi$. Introducing these values into (\ref{juan1}), (\ref{juan2}) and (\ref{juan3}), yields simplified expressions. Attending the isotropic case, Eq.~(\ref{type1gral}) acquires the form
\be
\lim_{\varepsilon_o \rightarrow \varepsilon_e} \left. d_{\operatorname{eff}}^{\operatorname{ooe}}\right\vert_{\operatorname{col}} = d_{31} \sin \sigma - d_{22} \cos \sigma \sin 3 \varphi,
\label{type1}
\ee
while both (\ref{eoe}) and (\ref{oee}) are reduced as follows 
\be
\lim_{\varepsilon_o \rightarrow \varepsilon_e} \left. d_{\operatorname{eff}}^{\operatorname{eoe}} \right\vert_{\operatorname{col}}  = \left. d_{\operatorname{eff}}^{\operatorname{oee}} \right\vert_{\operatorname{col}}  = d_{22} \cos^2 \sigma \cos 3 \varphi.
\label{type2}
\ee
Equations (\ref{type1}) and (\ref{type2}) were first reported in \cite{Mid65}, and then included in the reviews \cite{Nik91,Dmi99}. Here, they are recovered as particular cases of our results (\ref{juan1})-(\ref{oee}).

In general, the effective nonlinear coefficient $d_{\operatorname{eff}}$ is useful to calculate quantities such as conversion efficiency $\eta$, whose formulas often involve $d_{\operatorname{eff}}^2$ instead of the coefficient itself, see for instance Tables~2.28 and 2.29 in \cite{Dmi99}. 

Assuming that factors like group-velocity mismatch, dispersive spreading and diffraction can be neglected, the plane-wave fixed-field approximation yields expressions of $\eta$ that are proportional to the pump power density $P_p$, the square of the crystal length $L$, and the ``quality parameter'' $Q =d_{\operatorname{eff}}^2/(n_p n_s n_i)$, with $n_u$ the refractive index of the $u$-light wave~\cite{Dmi99} (an additional factor can be included to consider the effect of the wave mismatch on the conversion efficiency). However, convertible radiation is not a plane wave in real frequency converters, so the accurate calculation of $\eta$ is very complex. Analytic expressions of $\eta$ are feasible only for some special and simple cases. 

Next, we are going to analyze the situation in which a type-II BBO crystal is pumped by a violet laser diode that operates at $\lambda_p = 405$ nm. Configurations for other admissible values of the pump wave-length $\lambda_p$ (equivalently, the pump angular frequency $\omega_p$) are feasible from our general solutions.

\subsection{Degenerate case}

Throughout this section we assume that down-converted light is created in degenerate form ($\omega_s = \omega_i = \omega_p/2$). As noted above, in this case the Snell law (\ref{snell2}) leads to the identity $\theta_s^{(a)} = \theta_i^{(a)}$. Consequently, after refraction, the signal and idler cones have the same aperture and inclination with respect to the pump beam when $\omega_s = \omega_i$, as observed in the laboratory. 

We have already indicated that the roots $\chi_k$ introduced in (\ref{set}) are functions of $\phi_i$, $\sigma$, and the three angular frequencies $\omega_u$. Hereafter we use the shorthand notation $\chi_k = \chi_k (\phi_i, \sigma)$, $k=1,2,3,4$, with implicit dependence on the angular frequencies.

\subsubsection{Delimiting the orientation of the optic axis}

In general, the functions $\chi_k (\phi_i, \sigma)$ introduced in (\ref{set}) are complex-valued. However, only  real-valued functions $\chi_k$ can be directly associated with angles $\theta_i$ that are measured in laboratory. Therefore, as the condition $\phi_i \in  [0^\circ, 360^\circ)$ is mandatory, we have to differentiate the values of $\sigma$ that produce real-valued functions $\chi_k (\phi_i, \sigma)$ from those that give rise to $\chi_k (\phi_i, \sigma)$ with non-zero imaginary part.

\begin{figure}[h!]
\centering
\subfloat[][General roots]{\includegraphics[height=0.3\textwidth]{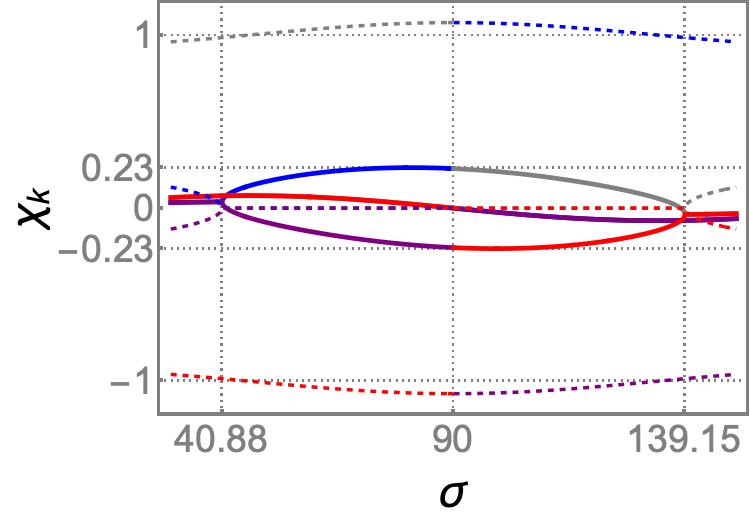}}
\hskip3ex
\subfloat[][Real-valued roots]{\includegraphics[height=0.3\textwidth]{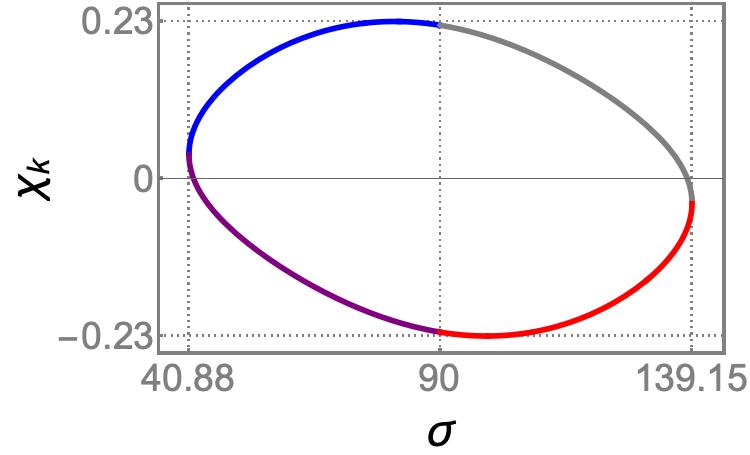}}

\caption{\footnotesize The functions $\chi_k (\phi_i, \sigma)$ introduced in Eq.~(\ref{set}) are complex-valued in general. Fixing $\phi_i = 0^\circ$, they have been plotted in blue, purple, gray, and red for $k=1,2,3,4$, with the real and imaginary parts represented by continuous and dotted curves (a). The real-valued versions are complementary in the interval $\Lambda_{\sigma} = [40.88^{\circ},139.15^{\circ}]$ (b). The plots consider down-converted light created in degenerate configuration ($\omega_i = \omega_s = \omega_p/2$), so they form an ellipse parameterized by $\sigma \in \Lambda_{\sigma}$, which is measured in sexagesimal degrees. 
}
\label{fig03}
\end{figure}

To get some insights on the matter consider the functions $\chi_k (\phi_i, \sigma)$ defined by $\phi_i = 0^\circ$. They are depicted in Figure~\ref{fig03} for $\sigma \in [0, 180^\circ]$, with a close-up to the vicinity where the imaginary part of  $\chi_k (0^\circ, \sigma)$ is equal to zero, Figure~\ref{fig03}(a). We identify two complementary intervals, $\sigma \in [40.88^{\circ}, 90^{\circ}]$ and $\sigma \in (90^{\circ},139.15^{\circ}]$, where $\chi_1, \chi_2 \in \mathbb R$ and $\chi_3, \chi_4 \in \mathbb R$, respectively. Other values of $\sigma$ yield $\chi_2 = \chi_1^*$ and $\chi_4 = \chi_3^*$, with $z^*$ the complex-conjugate of $z \in \mathbb Z$. Therefore, in the present case, the orientation $\sigma$ of the optic axis must be confined to the interval $\Lambda_{\sigma} = [40.88^{\circ}, 139.15^{\circ}]$, see Figure~\ref{fig03}(b).

In general, the dependence of $\theta_i$ on $\phi_i$ and $\sigma$ is obtained through the set $\chi_k (\phi_i, \sigma) \in \mathbb R$, after reversing Eq.~(\ref{new}). Following the shorthand notation introduced above we write $\theta_i = \theta_i (\phi_i, \sigma)$, remember that the dependence on the three angular frequencies $\omega_u$ is implicit. To facilitate the analysis of $ \theta_i (\phi_i, \sigma)$ let us consider in detail the situations defined by $\phi_i = \operatorname{const}$ first, and then by $\sigma = \operatorname{const}$.

\begin{figure}[h!]
\centering
\subfloat[][$\phi_i= 0^\circ, 22.5^\circ, 45^\circ,67.5^\circ$.] {\includegraphics[height=0.3\textwidth]{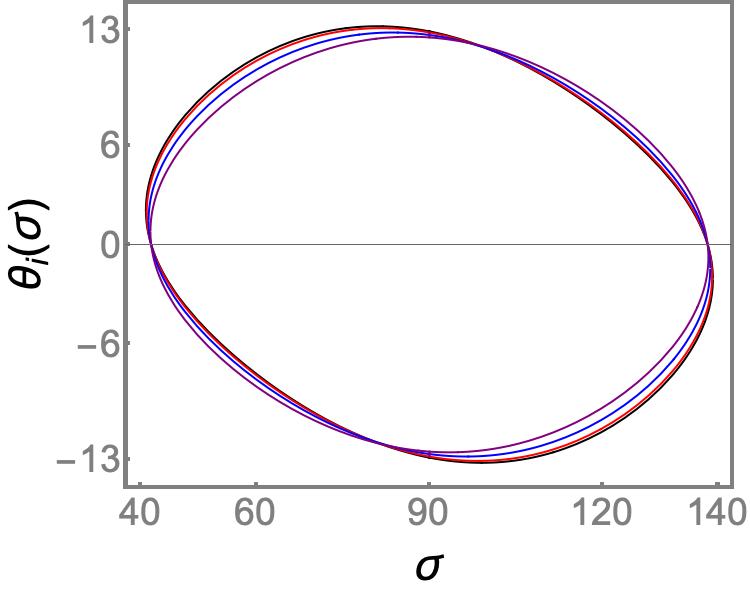}}
\hskip4ex
\subfloat[][$\phi_i = 0^\circ, 180^\circ$.] {\includegraphics[height=0.3\textwidth]{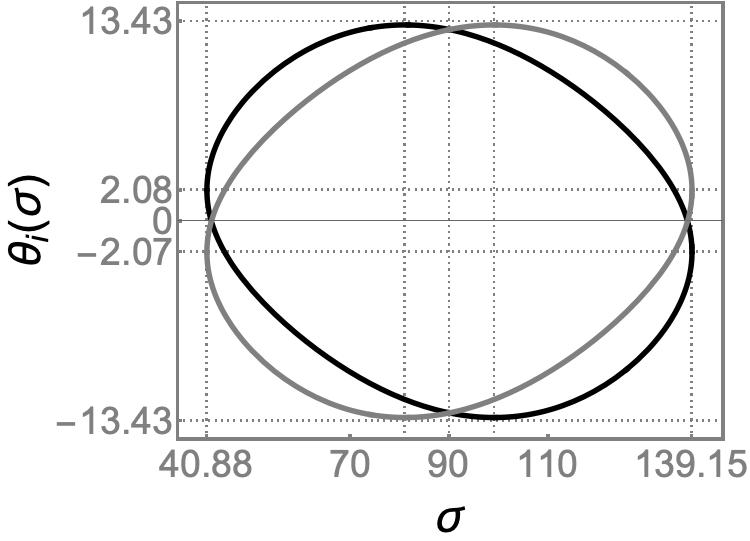}}

\caption{\footnotesize Given $\phi_i$, the polar angle $\theta_i(\sigma) \equiv \theta_i (\phi_i = \operatorname{const}, \sigma)$ describes an ellipse parameterized by $\sigma \in \Lambda_{\sigma}$. All the angles are measured in sexagesimal degrees. Black, red, blue, and purple paths stand for $\phi_i= 0^\circ, 22.5^\circ, 45^\circ$, and $67.5^\circ$ respectively (a). The paths of medium thickness in black and gray correspond to $\phi_i=0^\circ$ and $\phi_i=180^\circ$ (b). For $\phi_i =0^\circ$, the vertices are at the points $(40.88^\circ, 2.08^\circ)$ and $(139.15^\circ, -2.07^\circ)$, with maximum and minimum located at $(80.97^\circ, 13.43^\circ)$ and $(99.12^\circ, -13.43^\circ)$, and crossing points at $(41.79^\circ, 0^\circ)$ and $(138.2^\circ, 0^\circ)$. The ellipse defined by $\phi_i = 180^\circ$ is a counterclockwise rotated version of that generated by $\phi_i=0^\circ$. 
}
\label{fig04}
\end{figure}

$\bullet$ {\bf $\phi_i$ fixed.} The expression $\theta_i(\sigma) \equiv \theta_i (\phi_i = \operatorname{const}, \sigma)$ provides an ellipse parameterized by $\sigma \in \Lambda_{\sigma}$. Figure~\ref{fig04} shows the ellipses $\theta_i(\sigma)$ obtained for some representative values of the azimuthal angle $\phi_i$.

The points on the ellipse $\theta_i (\phi_i= \operatorname{const}, \sigma)$ are in a one-to-one relationship with the intersections of the cone of idler-light and the plane $\phi_i= \operatorname{const}$. The absence of intersections implies that there are no real angles $\theta_i$ derivable from the roots (\ref{set}) for the orientation $\sigma$, and vice versa. In the detection plane $\Sigma_0$, depending on the number of intersections, the condition $\phi_i= \operatorname{const}$ results in either a tangent or a secant to the transverse pattern formed by idler-light. The former case refers to a single intersection while the latter makes reference to two intersections. This identification is clearly associated with the fact that the ellipses $\theta_i(\sigma)$ provide at most two real angles $\theta_i$ for each $\sigma$ in $\Lambda_{\sigma}$.

Observe the symmetry between the ellipses for $\phi_i=0^\circ$ and $\phi_i = 180^\circ$ in Figure~\ref{fig04}(b). As they verify $\theta_i (\phi_i= 180^\circ, \sigma) = - \theta_i (\phi_i= 0^\circ, \sigma)$, it is clear that the entire set of points on a given ellipse $\theta_i (\sigma)$ is redundant. 

The intersections occurring on the positive semi-axis of $\phi_i=0^\circ$ are in one-to-one correspondence with those occurring on the negative semi-axis of $\phi_i = 180^\circ$, and vice versa. Since these lines are one the $\pi$-rotated version of the other, we realize that both of them provide twice the same information up to the phase $e^{i\pi}$. Similar conclusions are obtained for any other line $\phi_i= \operatorname{const}$ and its $\pi$-rotated version.

To eliminate redundancies we will take the convention of counting only the intersections associated with nonnegative values of $\theta_i$ for each line $\phi_i= \operatorname{const}$. In this way, the information provided by $\phi_i= \operatorname{const}$ and $\phi_i' = \pi + \operatorname{const}$ is complementary: the values $\theta_i<0$ that are not counted for $\phi_i= \operatorname{const}$ are recoverable from the values $\theta_i >0$ counted for $\phi_i'$ (after a change of sign). Our convention is consistent with the fact that polar angles are formally nonnegative. 

All the points on a given ellipse $\theta_i (\phi_i= \operatorname{const}, \sigma)$ constitute the exact solution of the problem for $\phi_i= \operatorname{const}$. With the convention introduced above, no one of these points is discarded. The complete determination of the polar angle of idler-light is obtained by exhausting all the values of $\phi_i \in [0^\circ, 360^\circ)$, which can be done by taking line $\phi_i=0^\circ$ and rotating it counterclockwise around the $z$-axis while counting intersections with the transverse pattern of idler-light, see Section~\ref{examples} for details.

The ellipses defined by $\phi_i =0^\circ$ and $\phi_i = 180^\circ$ are of major relevance since they refer to the intersections of idler-light and the $xz$-plane for each admissible value of $\sigma$. As the idler-cone is centered along the $x$-axis, the set $\Lambda_{\sigma}$ that we have delimited from such ellipses contains all the values of $\sigma$ for which the cone of idler-light exists. The ellipses defined by other values of $\phi_i$ are associated with different subsets of $\Lambda_{\sigma}$. In this sense, $\theta_i (0^\circ, \sigma)$ and $\theta_i (180^\circ, \sigma)$ serve as envelope of the set $\theta_i (\phi_i= \operatorname{const}, \sigma)$.

\begin{figure}[h!]
\centering
\includegraphics[height=0.26\textwidth]{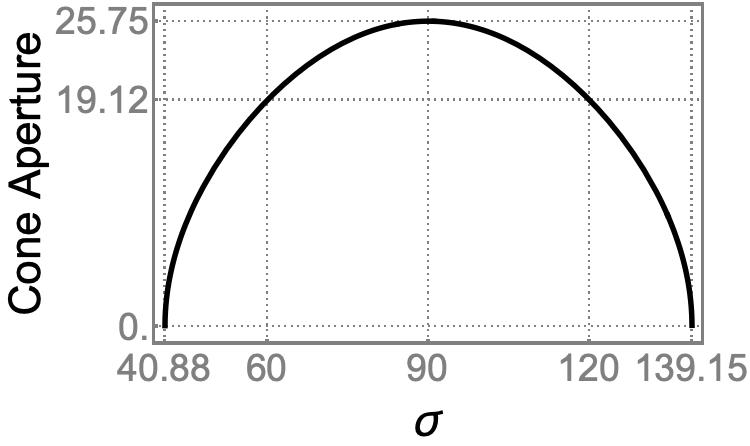}
\hskip4ex
\includegraphics[height=0.26\textwidth]{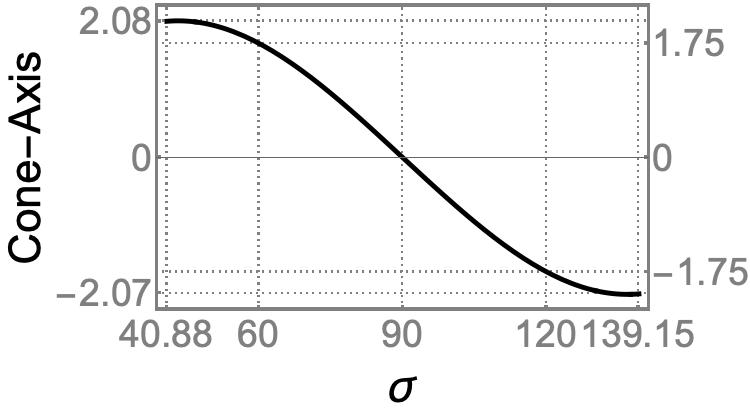}

\caption{\footnotesize Aperture $\Delta \theta_i$ and orientation of the cone-axis for idler light in terms of the orientation $\sigma$ of the optical axis. Both are expressed in sexagesimal degrees, and refer to the black-ellipse shown in Figure~\ref{fig04}(b).
}
\label{fig05}
\end{figure}

For $\phi_i = 0^\circ$, the difference $\Delta \theta_i$ of the corresponding polar angles provides the aperture of the cone. Figure~\ref{fig05}(a) shows $\Delta \theta_i$ as a function of $\sigma$. We can also characterize the orientation of the idler cone-axis in terms of $\sigma$. In Figure~\ref{fig05}(b) we appreciate that the cone-axis is located at the positive semi-axis $x>0$ for $\sigma < 90^\circ$, and it transits to $x<0$ for $\sigma >90^\circ$. At $\sigma = 90^\circ$, the cone-axis coincides with the pump-beam.

\begin{figure}[h!]
\centering
\includegraphics[height=0.3\textwidth]{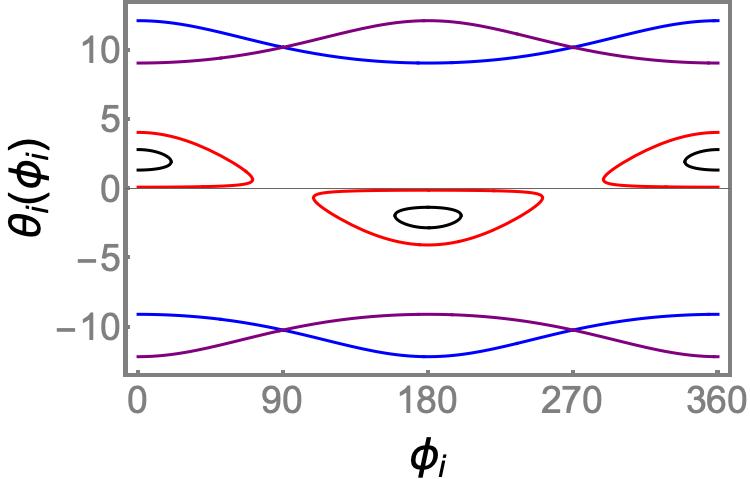}

\caption{\footnotesize Angular dependence for $\sigma = 41^\circ, 41.7^\circ$ (black and red lobes), and $\sigma =65^\circ, 115^\circ$ (blue and purple curves). The lobes are associated with cones of idler-light waves that describe circles entirely contained in either semi-plane $x \geq 0$ or $x\leq 0$ of the $xy$-plane. Continuous curves represent cones that describe circles occupying both semi-planes of the $xy$-plane.
}
\label{fig06}
\end{figure}

$\bullet$ {\bf $\sigma$ fixed.} The angular dependence $\theta_i(\phi_i)$ is illustrated in Figure~\ref{fig06} for representative values of $\sigma$ in the set $\Lambda_{\sigma}$. The presence of lobes (red and black paths in Figure~\ref{fig06}) means that $\theta_i(\phi_i)$ is real only for particular subsets of the azimuthal domain $\phi_i \in [0^\circ, 360^\circ)$ at the corresponding $\sigma$. Consistently, the transverse pattern of idler-light will be entirely contained in either semi-plane $x \geq 0$ or $x \leq 0$ of the detection plane $\Sigma_0$. In turn, the continuous curves shown in Figure~\ref{fig06} represent cones of idler-light whose transverse pattern occupies both semi-planes of $\Sigma_0$. Concrete configurations are discussed as examples in the next sections.

In order to correctly delimitate the angle $\sigma$, we have to impose the condition $\theta_i \in [0, \pi]$. Then, it is appropriate to consider the decomposition $\Lambda_{\sigma} = \Lambda_1 \cup \Lambda_2 \cup \Lambda_3$.

\begin{figure}[h!]
\centering
\includegraphics[height=0.26\textwidth]{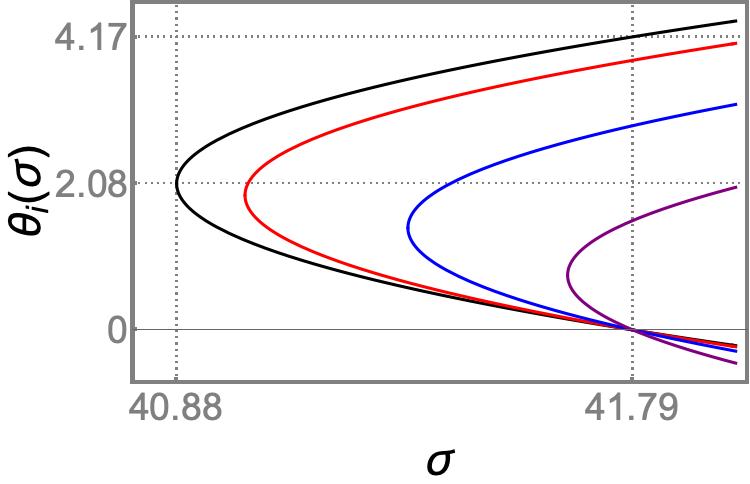}
\hskip6ex
\includegraphics[height=0.26\textwidth]{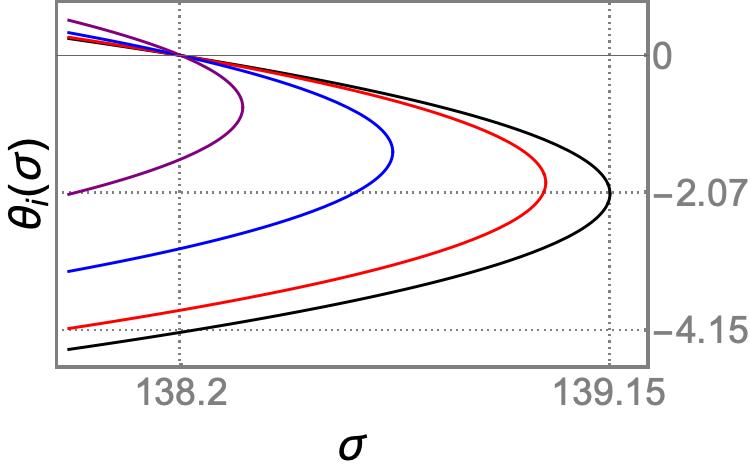}

\caption{\footnotesize  Details of the elliptic-like path shown in Figure~\ref{fig04}. The polar angle $\theta_i (\sigma) \geq 0$ is twice parameterized in the set $\Lambda_1 = [40.88^\circ, 41.79^\circ]$. The orientations $\sigma \in \Lambda_3= (138.2^\circ, 139.15^\circ]$ produce only negative values of $\theta_i (\sigma)$.
}
\label{fig07}
\end{figure}

According to our convention, the set $\Lambda_3 = (138.2^\circ, 139.15^\circ]$ is not taken into account for $\phi_i=0^\circ$ since $\sigma \in \Lambda_3$ yields only negative values of $\theta_i$, Figure~\ref{fig07}(b). In turn, from Figure~\ref{fig07}(a) we appreciate that $\Lambda_1 = [40.88^\circ, 41.79^\circ)$ provides two different values of $\theta_i >0$ for each value of $\sigma$, with exception of the vertices. This two-fold property of $\theta_i$ is associated with the lobes shown in Figure~\ref{fig06}. On the other hand, for $\Lambda_2 = [41.79^\circ, 138.2^\circ]$, the polar angle $\theta_i \geq 0$ can be operated as a function of $\sigma$. The latter is connected with the continuity of blue and purple curves of Figure~\ref{fig06}.

Therefore, for $\phi_i =0^\circ$ we formally constraint the orientations $\sigma$ of the optical axis to the set $\Lambda_{\sigma_*}= \Lambda_1 \cup \Lambda_2 = [40.88^\circ, 138.2^\circ]$. However, we have to keep in mind that the set $\Lambda_{\sigma}$ provides the orientations of the optical axis for which the idler cone exists either in $x>0$ or $x<0$.

\subsubsection{General configurations for down-converted light}
\label{hot}

Having in mind that our approach is based on exact solutions, the next qualitative profiles of the down-converted light hold for any admissible value of the pump wave-length $\lambda_p$, other than the violet one ($\lambda_p= 405 nm$) used as generic example throughout this work.

$\bullet$ For $\phi_i = 0^\circ$, the vertices of the ellipse  $\theta_i(\sigma)$ yield $\Delta \theta_i = 0^\circ$. These points correspond to the situation in which the cone structure of idler-light degenerates into a single beam. This phenomenon has been already described and measured in laboratory by other authors \cite{Kim05,Tak01,Kim03,Kwo08} but, as far as we know, its connection with the vertices of $\theta_i (\sigma)$ has been unnoticed in the literature. 

$\bullet$ At the crossing points with the $\sigma$-axis, the polar angle $\theta_i = 0^\circ$  defines a generatrix of the idler-cone that is collinear with the pump-beam. From Eq.~(\ref{ref0}) we realize that $\theta_i = 0^\circ$ implies $\theta_s = 0^\circ$, so there is also a generatrix of the signal-cone that is collinear with the pump-beam. The coincidence of these two generatrices with the pump-beam configures the down-converted cones in {\em osculating} form, where the cones touch along the common generatrix ($\theta_i = \theta_s = 0^\circ$) and are tangent to each other. 

From (\ref{kas}) we obtain $\mathbf k_i \vert_{\theta_i=0^\circ} =  k_i \hat{\mathbf e}_3$ and $\mathbf k_s \vert_{\theta_s=0^\circ} = k_s \hat{\mathbf e}_3$, so these wave-vectors, together with $ \mathbf k_p = k_p \hat{\mathbf e}_3$, satisfy the scalar (collinear) phase-matching. The latter has motivated the osculating configuration to be known as collinear \cite{Kie93,Pit95,Kwo08,Cas04}, but it should be emphasized that not all wave-vectors integrating the down-converted cones satisfy the scalar phase-matching. This is true only for wave-vectors along the generatrices that coincide with the pump beam, so the term ``collinear'' would be a misnomer for this configuration. 

$\bullet$ For $\sigma = 90^\circ$, the cone-axis of idler light coincides with the pump-beam while the cone aperture is maximum. The same holds for the signal-cone, so it overlaps exactly the idler-cone after refraction. In this case the down-converted cones are right-circular and, together with the pump beam, form the {\em coaxial configuration}. 

It is important to mention that $\sigma = 90^\circ$ reduces the fourth-order polynomial equation (\ref{monic}) to the bi-quadratic form discussed in Appendix~\ref{ApA}. Therefore, the coaxial configuration is the simplest and most symmetric profile of the down-converted cones derivable from the quartic equation (\ref{monic}).

Remarkably, for $\sigma = 90^\circ$, the authors of \cite{Kwi95b} indicate that ``no down conversion takes place at this setting'' since ``the actual down-conversion efficiency in a uniaxial crystal such as BBO varies as $\cos^2 (\theta_{pm})$ [\oldstylenums{22}]'', where $\theta_{pm}$ is the angle between the crystal optic axis and the pump beam (in our notation $\theta_{pm}= \sigma$), and [\oldstylenums{22}] corresponds to the first edition of our reference \cite{Dmi99}.

However, in the previous section we have pointed out that such statement is not entirely accurate. To be concrete, using (\ref{pump1}) and  (\ref{pump2}), the quality parameter for type-II BBO (e $\rightarrow$ e + o) acquires the form
\be
Q^{\operatorname{eoe}}  = \frac{\left( d_{\operatorname{eff}}^{\operatorname{eoe}} \right)^2}
{n_{pe} n_{po} n_{so} n_{ie} n_{io} } \left[ 
\left( n_{pe}^2 \cos^2 \sigma + n_{po}^2 \sin^2 \sigma \right)
\left( n_{ie}^2 \cos^2 \delta_i + n_{io}^2 \sin^2 \delta_i \right)
\right]^{1/2},
\label{quality}
\ee
recall that $n_{ue}$ and $n_{uo}$ are the extraordinary and ordinary refractive indexes for the $u$-wave. A similar expression is obtained for the other distribution of polarizations in type~II BBO (e $\rightarrow$ o + e).

Equation~(\ref{quality}) is consistent whenever the effective nonlinearity $d_{\operatorname{eff}}^{\operatorname{eoe}}$, originally calculated in terms of the angles $(\vartheta_u, \varphi_u)$ in the crystal, is expressed in terms of the laboratory angles $(\theta_u, \phi_u)$. The transformation is obtained through the rotation $R_y(-\sigma)$ of the wave-vectors $\mathbf k_u$, which gives rise to the relationships 
\bea
\cos \vartheta_u = \cos \sigma \cos \theta_u + \sin \sigma \sin \theta_u \cos \phi_u, 
\label{theta}\\[1ex]
\tan \varphi_u = \frac{ \sin \theta_u \sin \phi_u}{ \cos \sigma \sin \theta_u \cos \phi_u - \sin \sigma \cos \theta_u},
\label{phi}
\eea
where $\theta_p = \phi_p = 0^\circ$, so that $\vartheta_p= \sigma$ and $\varphi_p = 180^\circ$. Then, the most general effective nonlinearity (\ref{juan2}) is rewritten as follows
\be
\begin{array}{rl}
d_{\operatorname{eff}}^{\operatorname{eoe}} = 
& \left[ \left( \varepsilon_{pe}^2 \cos^2 \sigma + \varepsilon_{po}^2 \sin^2 \sigma \right) \left( \varepsilon_{ie}^2 \cos^2 \vartheta_i + \varepsilon_{io}^2 \sin^2 \vartheta_i \right) \right]^{-1/2} \times \\[1ex]
& \quad \left\{ \varepsilon_{pe} \left[- \varepsilon_{ie} d_{22} \cos \vartheta_i \cos (\varphi_i + \varphi_s ) + \varepsilon_{io}d_{31}  \sin \vartheta_i \sin \varphi_s   \right] \cos \sigma  
\right.\\[1ex] 
& \quad \quad \left.  +  \, \varepsilon_{po} \varepsilon_{io}d_{31} \cos \vartheta_i \sin (\varphi_i - \varphi_s)  \sin \sigma \right\},
\end{array}
\label{juan2b}
\ee
Using (\ref{theta}) and (\ref{phi}), the introduction of (\ref{juan2b}) into (\ref{quality}) shows an ellaborated dependence of $Q$ on $\sigma$ (remember, $\theta_i$ depends on $\sigma$ and $\phi_i$, while $\delta_i$ depends on $\theta_i$, $\sigma$, and $\phi_i$). Nevertheless, for $\sigma= 90^\circ$, the straightforward calculation yields
\be
\left. Q^{\operatorname{eoe}} \right\vert_{\sigma = 90^\circ} = \left. 
\frac {\left( n_{ie}^2 \cos^2 \delta_i + n_{io}^2 \sin^2 \delta_i \right)^{1/2} }{\varepsilon_{ie}^2 \cos^2 \delta_i + \varepsilon_{io}^2 \sin^2 \delta_i }
\sin^2 \theta_i \sin^2 \Theta (\theta_i, \theta_s, \phi_i)
\right\vert_{\sigma=90^\circ}
\frac{d_{31}^2 n_{ie}^3  \cos^2 \phi_i }{n_{pe} n_{so} n_{io} },
\label{Qeoe}
\ee
with $\cos \delta_i \vert_{\sigma = 90^\circ} = \sin \theta_i \cos \phi_i$, see Eq.~(\ref{del1}), and
\[
\Theta (\theta_i, \theta_s, \phi_i) = \arctan \left[ \tan \theta_i \sin \phi_i \right] + \arctan \left[ \tan \theta_s \sin \phi_i \right],
\]
where we have used (\ref{pm3}). The values of $\theta_i$ and $\theta_s$ at $\sigma = 90^\circ$ are derivable from our exact solutions, see for instance Figure~\ref{fig04}. 

As the azimuthal angle $\phi_i$ does not depend on $\sigma$, the expression provided in (\ref{Qeoe}) is indeed a function of $\phi_i$ and the angular frequencies $\omega_i$, $\omega_s$, and $\omega_p$. It reaches its maximum at $\phi_i = 0^\circ, 180^\circ$, and is equal to zero for $\phi_i = 90^\circ, 270^\circ$. 

That is, for vector phase-matching, with $\sigma = 90^\circ$, the production of down-converted light is not only different from zero but optimized along the $x$-axis.

Accordingly, the conversion efficiency $\eta$ is different from zero within the plane-wave fixed-field approximation, where $Q$ is included as a factor in the formula of $\eta$ \cite{Dmi99}. Other approaches leading to $\eta$ must also include $Q$ as a factor in such a way that the result is reduced to that of the plane-wave fixed-field approximation in the appropriate limit.

The situation changes for scalar phase-matching since the introduction of (\ref{type2}) into (\ref{quality}) leads to the expression
\be
\left. Q^{\operatorname{eoe}}_{\operatorname{col}} \right\vert_{\sigma = 90^\circ}   =
\left. Q^{\operatorname{eeo}}_{\operatorname{col}} \right\vert_{\sigma = 90^\circ}   =
\left.
\left( n_{ie}^2 \cos^2 \delta_i + n_{io}^2 \sin^2 \delta_i \right)^{1/2} 
 \cos^4 \sigma
\right\vert_{\sigma=90^\circ}
\frac{d_{22}^2}{n_{pe} n_{so} n_{ie} n_{io} }.
\label{bad}
\ee
Note that the quality parameter (\ref{bad}) describes only the idler and signal waves that have their wave-vector aligned with the pump-beam. In such a case, the situation is even more dramatic than anticipated in \cite{Kwi95b}, where the efficiency is announced to vary as $\cos^2 \sigma$. 

Nevertheless, above we have found that, for $\sigma = 90^\circ$, the down-converted light is produced in right-circular cones of maximum aperture, whose axes are aligned with the pump-beam. That is, no generatrix of any of the cones coincides with the pump-beam  if $\sigma= 90^\circ$, so neither $\mathbf k_i$ nor $\mathbf k_s$ is collinear with $\mathbf k_p$. In this sense, the quality parameter (\ref{bad}) shows that no down-converted light is produced such that its wave-vector is aligned with the pump beam for $\sigma = 90^\circ$, but it does not mean that down-converted light is not produced when the optical axis is tilted at $\sigma = 90^\circ$.

Therefore, unlike the negative statement made in Ref.~\cite{Kwi95b}, we have shown that the conversion efficiency of type-II BBO crystals is different from zero at $\sigma = 90^\circ$. Moreover, the production of down-converted light is optimized along the $x$-axis at the setting we are dealing with.

\subsubsection{Geometric distribution of down-converted light for $\lambda_p = 405$ nm}
\label{examples}

Let us discuss in detail the configurations mentioned in the previous section, together with two additional settings that are relevant in practice, for a pump-beam with $\lambda_p = 405nm$. They are classified according to  the values of the optical-axis orientation $\sigma$ as this sweeps $\Lambda_{\sigma_*} = \Lambda_1 \cup \Lambda_2 = [40.88^\circ, 138.2^\circ]$ from left to right.

$\bullet$ {\bf Beam-like configuration.} For $\sigma = 40.88^\circ$ one has $\Delta \theta_i =0^\circ$ at $\phi_i=0^\circ$, see Figures~\ref{fig05}(a) and \ref{fig07}(a), so the idler cone is deprived of structure and degenerates into a single beam that forms the angle $\theta_i = 2.08^\circ$ with the pump beam. As transverse pattern, the idler beam depicts a single spot on the positive semi-axis $x>0$ of $\Sigma_0$. Consistently, the transverse pattern of the signal cone is a single spot on the negative semi-axis $x<0$. These spots are equidistant from the origin. Our results are in complete agreement with theoretical and experimental works already published by other authors \cite{Kim05,Tak01,Kim03,Kwo08}. 

\begin{figure}[h!]

\begin{minipage}[b]{1\linewidth}
\centering
\subfloat[][divergent cones] 
{\includegraphics[height=0.25\textwidth]{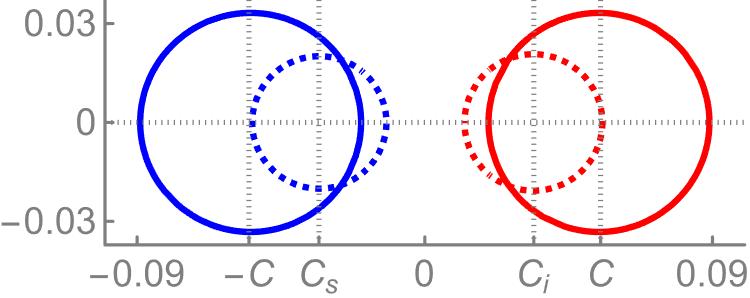}}
\end{minipage}
\vskip2ex
\centering
\begin{minipage}[b]{1\linewidth}
\centering
\subfloat[][idler angles] 
{\includegraphics[height=0.28\textwidth]{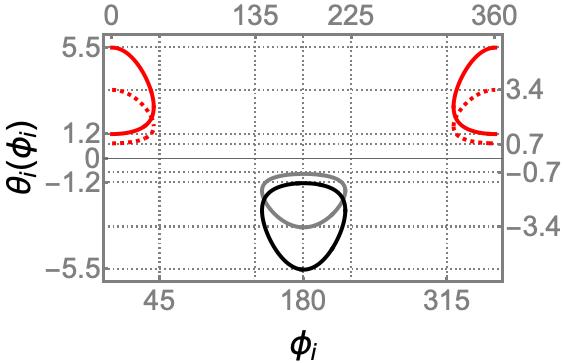}}
\hskip4ex
\subfloat[][signal angles] 
{\includegraphics[height=0.28\textwidth]{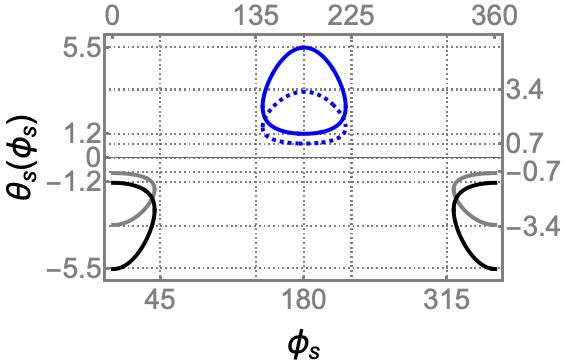}}
\end{minipage}

\caption{\footnotesize 
Configuration of the down-converted light for the optic-axis orientation $\sigma = 41.25^\circ$. The transverse pattern (a), depicted in the $xy$-plane for normalized wave-vectors $\hat {\mathbf k}_u = \mathbf k_u/k_u$, is formed by two separated rings that have their  centers along the $x$-axis. The polar angle $\theta$ is depicted against the azimuthal one $\phi$, both in sexagesimal degrees, for idler (b) and signal (c) beams. In all cases we distinguish between idler (red) and signal (blue) beams inside the crystal (dashed) and after refraction (continuous). The lobes in black (after refraction) and gray (inside the crystal) refer to negative values of the polar angles. They describe the behavior of the rings shown in (a) but in a system rotated by $\pi$ around the $z$-axis.
}
\label{fig08}
\end{figure}

$\bullet$ {\bf Divergent cones.} For $40.88^\circ < \sigma < 41.79^\circ$ the axes of idler cones diverge from the axis of the corresponding signal cone. The transverse pattern is formed by two separate rings that are centered along the $x$-axis; the idler-ring is entirely contained in the first and fourth quadrants of $\Sigma_0$, while the signal-ring is in the second and third quadrants. Inside the crystal, the centers are not equidistant from the origin $\vert C_i \vert \neq \vert C_s \vert$. After refraction (outside the crystal), $C_s=-C_i$, with $C_i = C>0$, so the axes of both cones are now  equidistant from the pump beam. Our results are in complete agreement with previous studies, see for instance \cite{Kwo08}. 

Figure~\ref{fig08} illustrates the case for $\sigma = 41.25^\circ$. Remark that the polar angle $\theta_i$ is positive in the vicinities of $\phi_i = 0^\circ$ (and $\phi_i = 360^\circ$), see Figure~\ref{fig08}(b). In turn, the polar angle $\theta_s$ of the signal beam is positive around $\phi_s =180^\circ$, see Figure~\ref{fig08}(c). For other values of the azimuthal angles, there is no real solution for the polar angles. 

The lobes described by $\theta_u$ as a function of $\phi_u$ are justified as follows. Let us measure the cone-aperture of idler light in terms of $\phi_i$, see the red rings in Figure~\ref{fig08}(a). Considering that $\phi_i$ runs counterclockwise, we can start by tracing a half-line along the positive semi-axis $x>0$ of $\Sigma_0$, with the left-edge fixed at origin. We find two intersections between the half-line and both red rings. These intersections correspond to the points on the left semi-lobe at $\phi_i=0^\circ$, see Figure~\ref{fig08}(b). Tilting the half-line towards the positive semi-axis $y>0$, the intersections collapse into a single one at $\phi_i \approx 45^\circ$, where the half-line is tangent to the ring and the left semi-lobe is completed. Increasing the value of $\phi_i$, the half-line finds nothing up to $\phi_i \approx 315^\circ$, where the lower part of the red rings is found (and the right semi-lobe starts). We exhaust the searching of intersections by tilting the half-line towards the positive semi-axis $x>0$. The blue lobe described by $\theta_s$ in Figure~\ref{fig08}(c) admits a similar description. 

With the previous description, it is clear that the negative values of the polar angles (lobes in black and dashed-gray) in Figures~\ref{fig08}(b) and \ref{fig08}(c) also describe the rings shown in Figure~\ref{fig08}(a), but in a reference system that is rotated by $\pi$ around the $z$-axis. This symmetry has been discussed above, where we have introduced the convention of considering only nonnegative values of $\theta_i$. The lobes in black and dashed-grey shown in Figure~\ref{fig08} reproduce $\theta_i >0$ in reverse order (with $\sigma$ sweeping $\lambda_{\sigma_*}$ from right-to-left), after a change of sign. In this form, we remark that all the real roots (negative and nonnegative) found in the previous sections are useful to describe the SPDC phenomena. 

\begin{figure}[h!]

\begin{minipage}[b]{1\linewidth}
\centering
\subfloat[][osculating cones] 
{\includegraphics[height=0.25\textwidth]{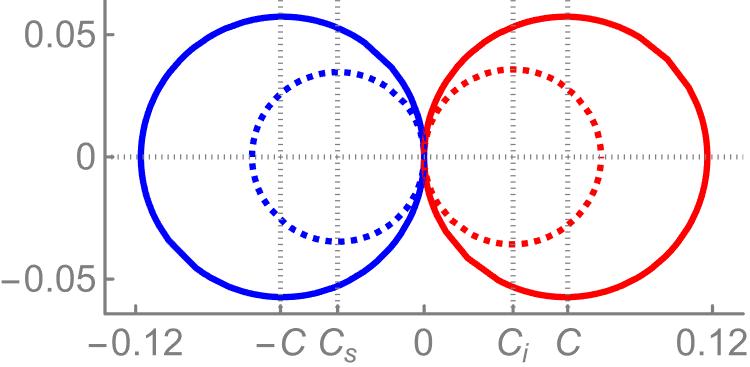}}
\end{minipage}
\vskip2ex
\centering
\begin{minipage}[b]{1\linewidth}
\centering
\subfloat[][idler angles] 
{\includegraphics[height=0.28\textwidth]{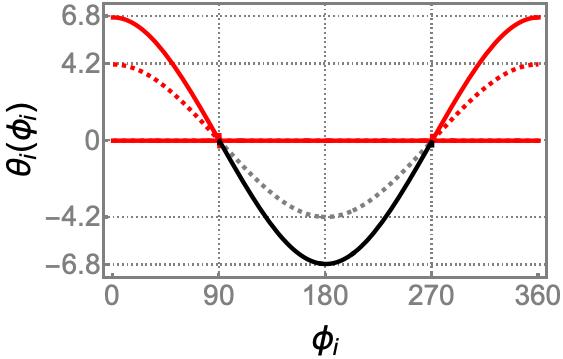}}
\hskip4ex
\subfloat[][signal angles] 
{\includegraphics[height=0.28\textwidth]{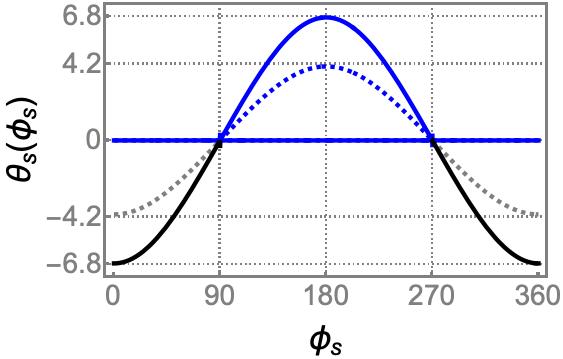}}
\end{minipage}

\caption{\footnotesize 
Configuration of the down-converted light for the optic-axis orientation $\sigma = 41.79^\circ$. The colors-code, units and nomenclature are the same as those indicated in Figure~\ref{fig08}.
}
\label{fig09}
\end{figure}

$\bullet$ {\bf Osculating cones.} For the right-edge of $\Lambda_1$, that is $\sigma = 41.79^\circ$, we obtain $\theta_i=0^\circ$, which defines a generatrix that is collinear with the pump-beam. This result is in complete agreement with the so called ``collinear configuration'' studied in references \cite{Kie93,Pit95,Kwo08,Cas04}, for instance.

The setting is illustrated in Figure~\ref{fig09}. Similar to the previous case, the cone-apertures $\Delta \theta_u$ are in correspondence with finite intervals of $\phi_u$, but now the down-converted light is produced in a pair of osculating cones.

As we have seen, positioning the crystal such that the angle formed by the optical axis and the pump beam is in the set $\Lambda_1 = [40.88^\circ, 41.79^\circ]$, we recover three different configurations of the down-converted light that are recurrently studied in the literature. 

It is important to emphasize that the width of $\Lambda_1$ is only $0.91^\circ$, so it seems challenging to align the optical axis with the appropriate precision to distinguish between the configurations associated with the extremes of $\Lambda_1$, and that linked with any other $\sigma \in \Lambda_1$. Namely, great precision is required to achieve the beam-like ($\sigma = 40.88^\circ$), divergent ($40.88^\circ < \sigma < 41.79^\circ$), and osculating ($\sigma = 41.79^\circ$) configurations in the laboratory. Fortunately, in practice this is not a major problem. For instance, in \cite{Kwo08} it is reported a series of experimental studies where beam-like, osculating (collinear), and ``non-collinear'' configurations are compared. The pump laser was a 408 nm cw diode laser, and the SPDC photons had the central wavelength of 816 nm. For the osculating configuration it is reported the angle $\sigma = 41.5^\circ$ (in our notation), while the beam-like was accomplished by reducing the angle to $\sigma = 40.6^\circ$. These values of $\sigma$ define an interval of width $0.9^\circ$, which is very close to the width of $\Lambda_1$ mentioned above. That is, it is feasible to achieve the precision required to experimentally produce the configurations we are dealing with.

Remember that although we are discussing the consequences of pumping light at $\lambda_p = 405$ nm into a BBO crystal, our exact solutions allow to consider any admissible value of $\lambda_p$, in particular $\lambda_p = 408$ nm as this was used in \cite{Kwo08}. Concrete examples are discussed below.  

In addition to the previous configurations, two additional settings are feasible if $\sigma \in (41.79^\circ, 90^\circ] \subset \Lambda_2$. They are as follows:

\begin{figure}[h!]
\centering
\subfloat[][Overlapping cones ($\sigma=60^\circ$)] {\includegraphics[height=0.30\textwidth]{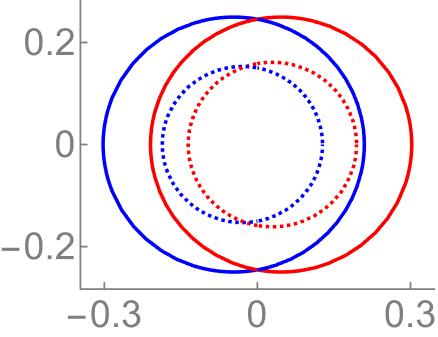}}
\hskip4ex
\subfloat[][Coaxial cones ($\sigma=90^\circ$)]{\includegraphics[height=0.30\textwidth]{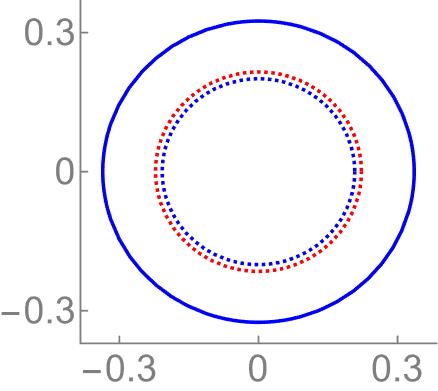}}

\caption{\footnotesize  Configurations of the down-converted light for the optic-axis orientations $\sigma = 60^\circ$ (a) and $\sigma = 90^\circ$ (b). The colors-code, units and nomenclature are the same as those indicated in Figure~\ref{fig08}.
}
\label{fig10}
\end{figure}

$\bullet$ {\bf Overlapping cones.} Given $41.79^\circ < \sigma < 90^\circ$, the polar angle $\theta_u$ is different from zero for any value of $\phi_u \in [0, 2\pi)$ in both down-converted cones. The latter means that the cones overlap by preserving their centers along the $x$-axis. The larger the value of $\sigma$, the shorter the distance between the centers. 

The overlapping configuration is illustrated in Figure~\ref{fig10}(a) for $\sigma = 60^\circ$, it is also classified as non-collinear, see for instance \cite{Kwi95b,Kwo08}. The points of intersection deserve special attention since they identify maximally entangled states \cite{Wal10,Cou18,Zha21,Rub94,Rub96,Kwi95b}.

$\bullet$ {\bf Coaxial configuration.} For $\sigma =90^\circ$, the axes of the down-converted cones are along the $z$-axis. Thus, the axes of idler and signal cones are aligned with the pump-beam, so these three waves of light form the coaxial structure shown in Figure~\ref{fig10}(b). 

To conclude this section notice that the complementary set $\sigma \in (90^\circ, 138.2^\circ] \subset \Lambda_2$ reproduces the overlapping and osculating configurations in reverse order: the larger the value of $\sigma$, the longer the distance between the centers. In this sense, we can take $\Lambda_* = [40.88^\circ, 90^\circ] \subset \Lambda_{\sigma_*}$ as the definite set of optic-axis orientations producing type-II SPDC in the different configurations discussed above.

\subsection{Non-degenerate case}
\label{ndc}

The exact solutions $\chi_k$ introduced in Eq.~(\ref{set}) can be used with the down-converted frequencies rewritten as $\omega_i = \kappa \omega_p$ and $\omega_s = (1-\kappa) \omega_p$. To get nonzero frequencies within the frequency-matching regime, the parameter $\kappa$ is restricted to the interval $(0 ,1)$. Using this notation we also write $\lambda_i = \lambda_p/\kappa$ and $\lambda_s = \lambda_p/(1-\kappa)$ for the related wavelengths. 

Degenerate frequencies $\omega_i = \omega_s = 0.5 \omega_p$ are recovered at $\kappa =0.5$. For $\kappa \neq 0.5$, we have non-degenerate frequencies $\omega_i \neq \omega_s$ fulfilling frequency-matching. However, very low frequencies arise at $\kappa \rightarrow 0$ or $\kappa \rightarrow 1$, so excessively long wavelengths would be calculated for down-converted light in a given crystal. 

The latter shows that not all possible combinations of $\omega_i$ and $\omega_s$  that add up to $\omega_p$ lead to appropriate results. Therefore, we must further restrict the values of $\kappa$. The key to refine this parameter is provided by the intrinsic properties of the crystal under study. In fact, uniaxial nonlinear crystals operate in a very specific range of wavelengths that is defined by Sellmeier equations. We will take full advantage of this property to define $\kappa$.

The down-converted frequencies that obey both the frequency-matching and the restrictions due to the intrinsic properties of a given crystal define the variability of the non-degenerate case. In general, the configurations calculated for converted light of degenerate frequencies are affected when $\omega_i \neq \omega_s$, both in the aperture of the cones and in the arrangement of their axes. The greater the difference between $\omega_i$ and $\omega_s$, the more noticeable the changes. 

One of the reasons for studying non-degenerate frequencies is that real crystals are imperfect. We know that even when they are designed to produce converted light at degenerate frequencies from a pump frequency $\omega_p$, the frequencies of the resulting light are centered at $0.5 \omega_p$ and sweep over a range of values whose width is usually inversely proportional to the quality of the crystal, but is never equal to zero. 

Assuming that the bandwidth $\Delta \omega = \omega_p \Delta \kappa$ of the converted frequencies is defined by $0 < \Delta \kappa \ll 0.5$, it is immediate to see that the conical surfaces of converted light are not indefinitely thin, but have a structure whose width is determined by $\Delta \kappa$. The transverse patterns observed in the detection plane $\Sigma_0$ are actually rings whose width is also determined by $\Delta \kappa$. Therefore, in order to have a realistic description of down-conversion, our model also takes into account these variations from the ideal case.

To investigate the properties of the non-degenerate case we will focus on BBO crystals and pump beams at $\lambda_p = 405$ nm, with which we have been working on as an example. However, we must insist that our theoretical model is exact and general, so it is useful to study any other uniaxial crystal and other pump wavelengths.

\begin{figure}[h!]
\centering
\includegraphics[height=0.3\textwidth]{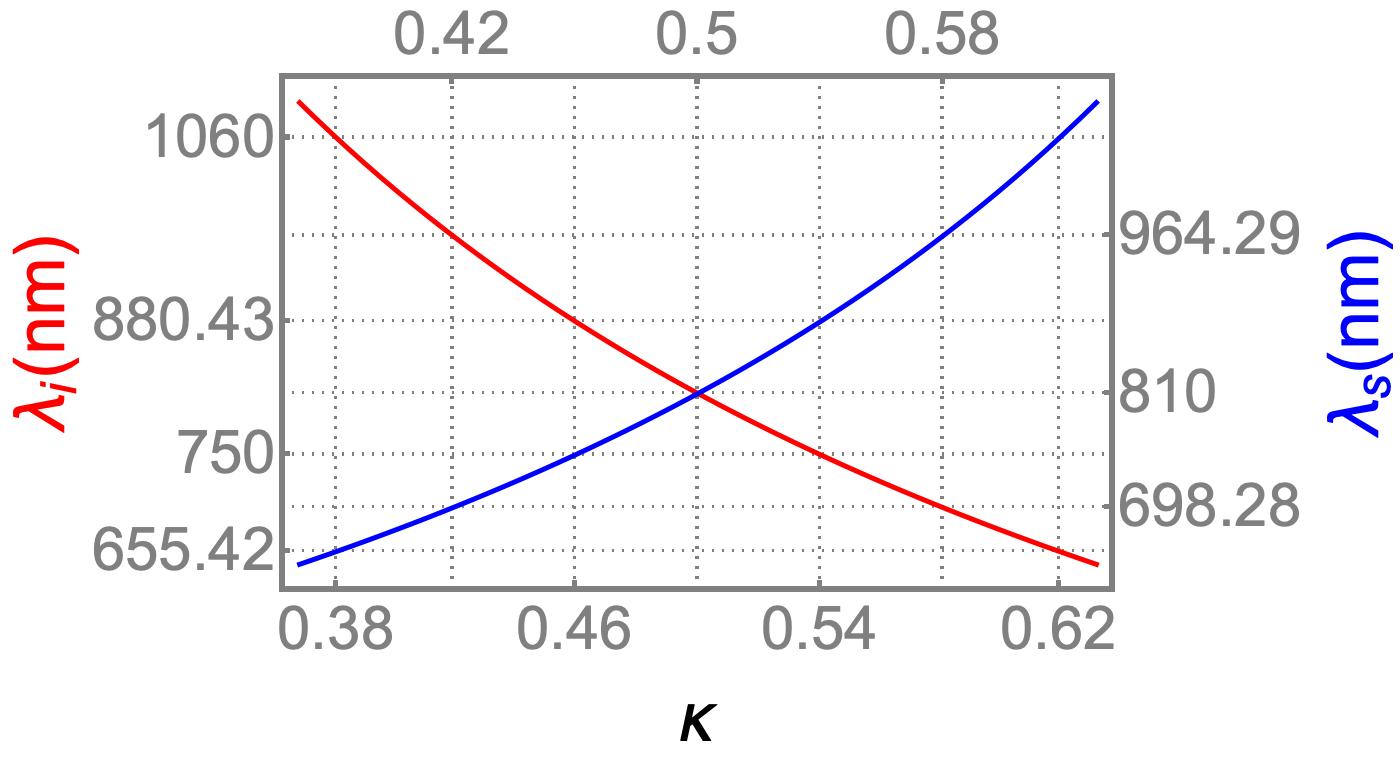}

\caption{\footnotesize Permissible wavelengths for down-converted light in a BBO crystal that is pumped at $\lambda_p = 405$ nm. For $\kappa \in (0.382,0.6179)$, the wavelengths $\lambda_i = \lambda_p/\kappa$ and $\lambda_s = \lambda_p/(1-\kappa)$ are in the range (655.4198 -- 1060) nm, in agreement with the Sellmeier equations (\ref{uno})-(\ref{dos}). Equivalently, the permissible down-converted frequencies $\omega_i = \kappa \omega_p$ and $\omega_s = (1-\kappa) \omega_p$ are in the range (0.382  -- 0.6179) $\omega_p$.
}
\label{fig11}
\end{figure}

For BBO crystals, the Sellmeier equations (\ref{uno})-(\ref{dos}) are valid for $\lambda$ in the range (220 -- 1060) nm. Demanding the wavelengths $\lambda_i$ and $\lambda_s$ to be in such a range we obtain $\kappa \in (0.382,0.6179)$. Therefore, the range of permissible wavelengths for down-converted light is  (655.4198 -- 1060) nm, the results are shown in Figure~\ref{fig11}. Equivalently, the permissible down-converted frequencies are in the range  (0.382  -- 0.6179)$\omega_p$.

With $\omega_i$ and $\omega_s$ in their permissible range, the polar angle $\theta_i$ presents some particularities that are not evident when looking only at the degenerate value $\omega_i = \omega_s$. In particular, one finds a close relationship between the optical axis orientation and the converted frequencies that affects the production of converted light, even to the extent of canceling the down-conversion. This accounts for the sensitivity to alignment between the optical axis and the pump beam found in the laboratory when producing down-converted light.

\begin{figure}[h!]
\centering
\includegraphics[height=0.3\textwidth]{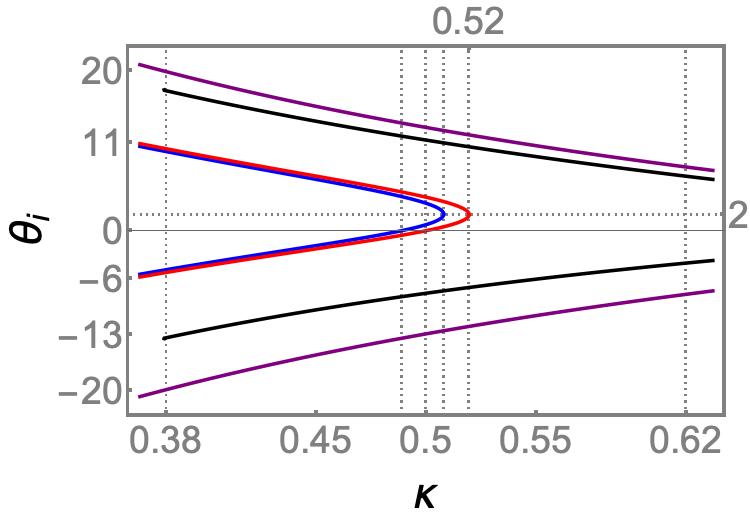}

\caption{\footnotesize Ellipses $\theta_i (\phi_i = 0^\circ, \sigma) \equiv \theta_i(\sigma)$ in the range of permissible down-converted wavelengths shown in Figure~\ref{fig11}. Curves in blue, red, black, and purple  refer to $\sigma = 41.25^\circ$, $41.79^\circ$, $60^\circ$, and $90^\circ$, respectively. At $\kappa = 0.5$, they provide the polar angle $\theta_i$ in the $xz$-plane of the laboratory frame for divergent, osculating, overlapping, and coaxial configurations. The vertices of blue and red ellipses are located respectively at $(\kappa, \theta_i) =(0.5078, 2.0559^\circ)$ and $(0.5193, 2.0148^\circ)$. These latter curves intersect the $\sigma$-axis at $\kappa =0.5$  and $\kappa = 0.4888$.
}
\label{fig12}
\end{figure}

Figure~\ref{fig12} shows the ellipses $\theta_i(\phi_i= 0^\circ, \sigma)$ in the range of permissible down-converted frequencies derived above. To compare with the results presented in the previous section, we have taken $\sigma = 41.25^\circ$ (divergent), $41.79^\circ$ (osculating), $60^\circ$ (overlapping), and $90^\circ$ (coaxial). In this form, the plots shown in Figure~\ref{fig12} represent the polar angle $\theta_i$ in the $xz$-plane of the laboratory frame, as a function of the non-degenerate variability $\kappa$. The results found in the previous section for $\phi_i =  0^\circ$ are recovered when $\kappa =0.5$. The ellipses $\theta_i(\phi_i= \operatorname{const}, \sigma)$ for other values of the azimuthal angle $\phi_i$ behave in similar form.

The behavior of the blue and red ellipses is particularly striking since they run out almost as soon as $\omega_i > 0.5 \omega_p$. This means that converted light occurs at most with frequencies $\omega_i$ slightly above the degenerate frequency when the optic axis is tilted at $\sigma =41.25^\circ$ or $\sigma =41.79^\circ$. Let us analyze these cases separately.

Assuming the optic axis is at $\sigma = 41.25^\circ$ (blue curve), from left to right in Figure~\ref{fig12}, the intersection of the ellipse and the $\sigma$-axis occurs at $\omega_{i, 1} = 0.4888 \omega_p$, while the vertex is defined by $\omega_{i,2} = 0.5078 \omega_p$. The latter is the upper bound of permitted frequencies $\omega_i$ in this case. The idler-light transforms from an osculating to a divergent configuration, and then to a beam-like configuration as $\omega_i$ goes from $\omega_{i,1}$ to the degenerate value, and then to the maximum value $\omega_{i,2}$. 

If all the frequencies between $\omega_{i,1}$ and $\omega_{i, 2}$ are in the bandwidth $\Delta \omega$, then the above configurations could not be distinguished from each other. For narrow enough bandwidths, some or all of these configurations could be studied separately. Therefore, the transverse pattern (a disk or a ring) will depend primarily on how narrow is the bandwidth of a given crystal. The only exception is the beam-like configuration, since its transverse pattern always forms a disk. Of course, other factors include the technical ability to align the optical axis with respect to the pump beam.

Equivalently, tilting the optic axis to $\sigma =41.79^\circ$, the key frequencies are the degenerate and $\omega_{i,3} =  0.5193 \omega_p$, which is now the upper limit. The transformation of the idler-light settings is quite similar to the previous case, but now from $0.5 \omega_p$ to $\omega_{i,3}$. The discussion about distinguishability of the configurations is the same. 

As we can see,  non-degenerate frequency-matching includes some subtleties that defy experimental skills. High precision is required to measure the orientation of the optic axis, as well as crystals characterized by a very fine bandwidth to distinguish between osculating, diverging, and beam-like configurations.

The sensitivity of divergent and osculating configurations is not present in overlapping and coaxial settings, where deviations from the degenerate frequency involve only smooth variations of $\theta_i$, see curves in black and purple in Figure~\ref{fig12}. This stability allows a better investigation of the behavior of down-converted light beyond the ranges that characterize the imperfection of crystals.

\begin{figure}[h!]
\centering
\subfloat[][$\omega_i=0.54 \omega_p$] {\includegraphics[height=0.25\textwidth]{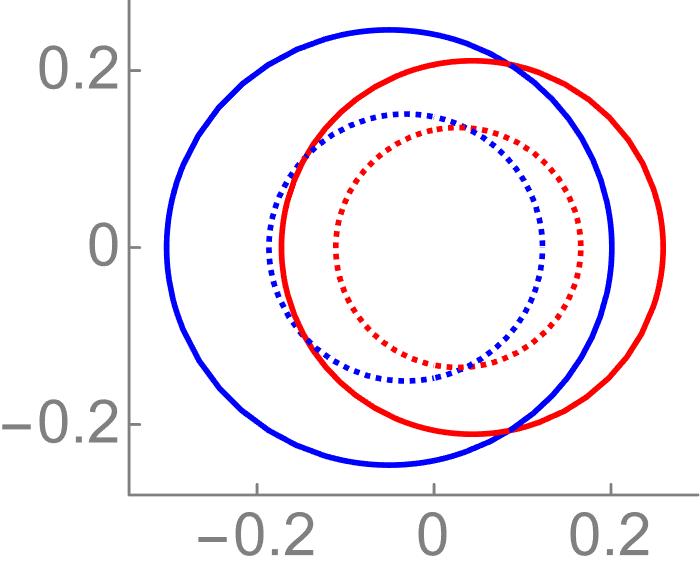}}
\hskip4ex
\subfloat[][$\omega_i=0.58 \omega_p$]{\includegraphics[height=0.25\textwidth]{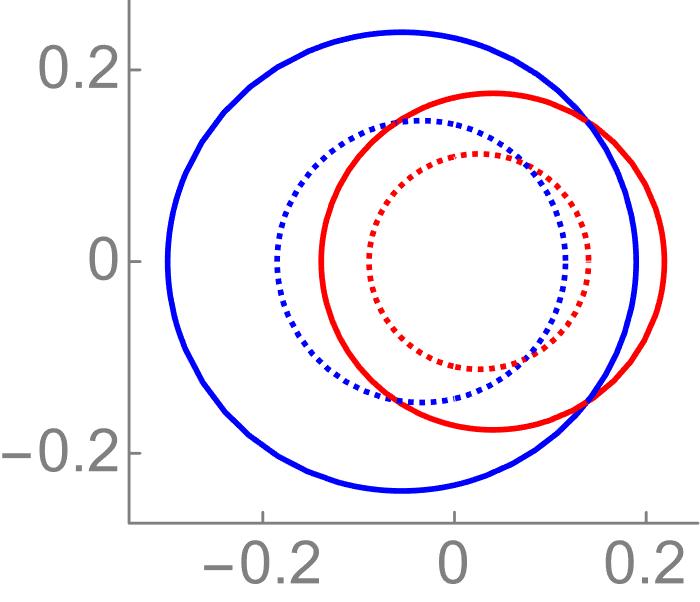}}
\hskip4ex
\subfloat[][$\omega_i=0.6179 \omega_p$]{\includegraphics[height=0.25\textwidth]{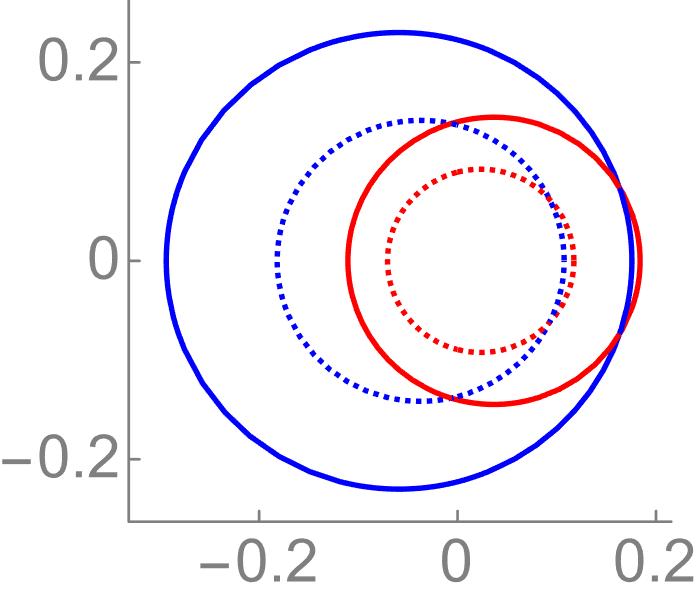}}

\caption{\footnotesize Configuration of the down-converted light for $\sigma = 60^\circ$ in the non-degenerate case. Frequency-matching $\omega_p = \omega_i + \omega_s$ is satisfied for the indicated values of $\omega_i$. The colors-code, units and nomenclature are the same as those indicated in Figure~\ref{fig08}; compare with Figure~\ref{fig10}(a).
}
\label{fig13}
\end{figure}

Figure~\ref{fig13} shows the geometric distribution of down-converted cones for $\sigma= 60^\circ$ and three different values of $\omega_i \neq 0.5 \omega_p$. Compared with the case of degenerate frequencies shown in Figure~\ref{fig10}(a), the cones do not have the same aperture. In fact, when $\omega_i$ is increased with respect to the degenerate value, the centers of the circles get closer while the aperture of idler-cone decreases. Consequently, the intersection points define a segment of line that is parallel but not coincident with the $z$-axis, and the distance between such points is shortened.

From Figure~\ref{fig11}, the wavelengths of down-converted light in Figure~\ref{fig13}(a) are $\lambda_i = 750$ nm and $\lambda_s = 880.43$ nm. Similarly, in Figure~\ref{fig13}(b) they are $\lambda_i = 698.28$ nm and $\lambda_s = 964.29$ nm, while Figure~\ref{fig13}(c) refers to $\lambda_i = 655.42$ nm and $\lambda_s = 1060$ nm. 

According to our previous discussion, each of the above configurations could be at the center of a band width, so its transverse pattern would form a ring. If by any chance the frequencies of the three configurations shown in Figure~\ref{fig13} are in the same bandwidth, then these configurations are part of the same ring. The same conclusions hold for the cones of signal light (blue circles) shown in the figure.

\section{Discussion of results and conclusions}
\label{sec4}

Throughout the previous sections, we have taken $\lambda_p = 405$ nm as the prototypical wavelength of a pump beam that is injected into a BBO crystal to produce down-converted light. Under degenerate frequency-matching, down-converted light occurs at $\lambda_i = \lambda_s = 810$ nm, as is well known. In practice, such a light pump is generated by a violet laser while the down-converted photons are collected on infrared photo-detectors somewhere in front of the crystal. 

However, $\lambda_p = 405$ nm is not the only useful wavelength for this purpose. Other commonly used wavelengths are, for example, $\lambda_p = 402$ nm and $\lambda_p= 408$ nm. Our theoretical model leads to correct results for these and other wavelengths that produce parametric down-conversion in nonlinear uniaxial crystals in general, and in BBO crystals in particular.

\begin{figure}[h!]
\centering
\includegraphics[height=0.3\textwidth]{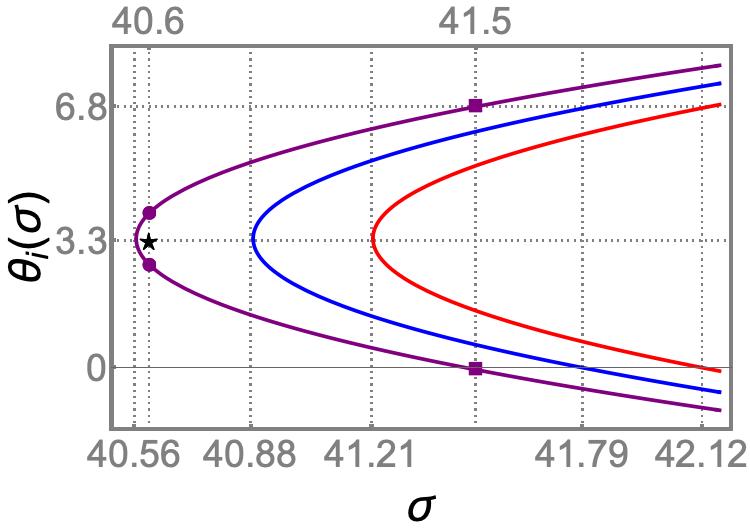}

\caption{\footnotesize  Ellipses $\theta_i( \phi_i= 0^\circ, \sigma)$ for $\lambda_p=$ 402, 405, and 408 (nm); respectively red, blue, and purple curves. Experimental measurements (black star) determined beam-like behavior at $(\sigma, \theta_i) = (40.6^\circ, 3.3^\circ)$ for $\lambda_p = 408$ nm \cite{Kwo08}. Our theoretical construction gives $(\sigma, \theta_i)=(40.5636^\circ, 3.3528^\circ)$, so the difference between experimental data and theoretical results is $(\delta \sigma, \delta \theta_i) = (0.0364^\circ, -0.0528^\circ$). In turn, the osculating configuration reported in \cite{Kwo08} was observed at $(41.5^\circ, 0^\circ)$, which contrasted with our theoretical result $(41.4711^\circ, 0^\circ)$ yields $\delta \sigma = 0.0289^\circ$. Based on the optic axis orientations measured in \cite{Kwo08}, the markers on the purple-ellipse determine the cone apertures $\Delta \theta_i = 1.3489^\circ$ (filled circles) and $\Delta \theta_i = 6.8459^\circ$ (filled squares).
}
\label{fig14}
\end{figure}

Figure~\ref{fig14} shows the polar angle of idler-light for $\lambda_p = 402, 405, 408$ nm, at $\phi_i=0^\circ$ (that is, in the $xz$-plane), as a function of the optic-axis orientation $\sigma$, and fulfilling the frequency-matching in degenerate form. We can see that the beam-like configuration (left vertices) is achieved at different orientations of the optical axis for different pump wavelengths. The same occurs for osculating configuration (intersections of the ellipses and the $\sigma$-axis). 

According to the notation introduced in the previous sections, we can write $\Lambda_1$ for the set of orientations of the optic axis that give rise to beam-like configuration at one extreme and osculating configuration at the other, with the divergent configuration in between. 

A striking feature of the results shown in Figure~\ref{fig14} is that the width of $\Lambda_1$ is around $0.9^\circ$, no matter the value of $\lambda_p$
(actually gives $0.9075^\circ$ for the blue and purple curves, and $0.9074^\circ$ for the red curve). It seems to us that this quantity so specific for the width of $\Lambda_1$ is a characteristic of the crystal.

One way to verify our conjecture could be to measure the width of $\Lambda_1$ for each of the wavelengths in Figure~\ref{fig14} (and as many others as possible), and find that, in fact, the same result is obtained regardless of the value of $\lambda_p$. 

Remarkably, the experimental measurements reported in \cite{Kwo08}, performed on a BBO crystal injected with a pump laser at $408$ nm, give exactly the width $0.9^\circ$ for $\Lambda_1$. This close agreement with our theoretical model represents a first step to verify the conjecture about the invariance of the width of $\Lambda_1$ with respect to pump wavelength.

However, as far as we have been able to review, the available data regarding $\lambda_p = 402$ nm and $\lambda_p = 405$ nm are not useful to complete the verification of the above conjecture, since the relationship of the reported results with the optic-axis orientation $\sigma$ is not explicitly provided. 

Experimental measurements are outside the scope of this work. Therefore, the complete experimental corroboration of the width invariance of $\Lambda_1$ with respect to $\lambda_p$, and its adjudication as a property of BBO crystals, remains an open question.

The other experimental results of \cite{Kwo08} are also in excellent agreement with our model. The beam-like and osculating configurations of converted light were observed at $\sigma =40.6^\circ$ and $41.5^\circ$, respectively. The difference with our theoretical result is $\delta \sigma = 0.0364^\circ$ and $\delta \sigma = 0.0289^\circ$, see details in Figure~\ref{fig14}. In turn, the overlapping configuration, accomplished at $\sigma = 42.3^\circ$ \cite{Kwo08}, is also in agreement with our model (although not included in Figure~\ref{fig14}).

Another noteworthy aspect of the information shown in Figure~\ref{fig14} is that the deviation $\delta \sigma = 0.0364^\circ$ of the experimental optic axis orientation ($\sigma = 40.6^\circ$) from the theoretical result ($\sigma = 40.5636^\circ$) implies different points of view about the configuration involved. Where the experimental measurements in \cite{Kwo08} yield a beam-like configuration, the theoretical result predicts a cone for the idler-light. 

In fact, at $\sigma = 40.6^\circ$ the purple ellipse provides $\theta_i = 4.02805^\circ$ and $\theta_i = 2.67908^\circ$, see the filled circles in Figure~\ref{fig14}. Then, we obtain the cone aperture $\Delta \theta_i = 1.34897^\circ$. In other words, tilting the optic axis at $\sigma = 40.6^\circ$ the cone of idler-light has not yet degenerated into a single beam, it still has structure!

Therefore, although the black star (experimental data) and the vertex of purple curve (theoretical result) are in good agreement, we wonder about the structure of the spots reported in Figure~1(c) of \cite{Kwo08}. Perhaps by placing the detection plane at a greater transverse distance from the crystal, an incipient ring structure could be observed.

In any case, as we have already discussed, the sensitivity of both theoretical and experimental results to small variations in $\sigma$ for divergent and beam-like configurations is extremely remarkable. 

Sensitivity to parameter variations is also seen in non-degenerate frequency-matching, where small deviations from the degenerate frequency could imply going abruptly from a given configuration to beam-like or overlapping configurations. This would even mean stopping the related emission of converted photons, as explained in Section~\ref{ndc}.

An unexpected result of our theoretical model is the prediction of right-circular cones for down-converted light at $\sigma = 90^\circ$. The aperture of both cones is maximum and their axes coincide with the pump-beam, so they overlap exactly to establish the configuration that we have called coaxial (Section~\ref{hot}). We have analyzed the conditions for the production of down-converted light in this configuration. Our results suggest that this configuration is feasible and that it is optimized along the $x$-axis for vector phase-matching. The situation changes for scalar phase-matching since no down-converted light is produced such that its wave-vector is aligned with the pump beam for $\sigma = 90^\circ$. However, the latter is completely consistent with the right-circular cones described above. 

We would like to emphasize that our model is based on the exact solutions of  vector phase-matching  for nonlinear uniaxial crystals. It is known that the corresponding equations are strongly transcendental since the  refractive indexes of extraordinarily polarized light are very elaborated functions of the unknowns, and the latter are encapsulated by trigonometric functions. Over the years, this fact has motivated more the study of numerical approaches than the search for analytical solutions. However, we have shown that the complexity of solving the strongly transcendental equations of vector phase-matching is reduced by transforming them into a fourth-order polynomial equation.

All the results reported in this work have been obtained by requiring a reality condition for the four roots of the quartic equation, since these are complex-valued in general. Such a condition defines a natural way to identify the orientations of the optical axis that are useful to produce down-conversion in the uniaxial crystal under study.

As we have seen, the model is in good agreement with available experimental data. Theoretical predictions like the invariance of the width of $\Lambda_1$ under the change of the pump wavelength, modifications to the configurations of down-converted light due to non-degenerate frequency-matching, or the production of down-converted light forming right-circular cones for $\sigma = 90^\circ$, await for experimental verification.

\appendix
\section{The fourth-order polynomial equation for type~II spontaneous parametric down conversion}
\label{ApA}

\renewcommand{\thesection}{A-\arabic{section}}
\setcounter{section}{0}  

\renewcommand{\theequation}{A-\arabic{equation}}
\setcounter{equation}{0}  

The quartic equation (\ref{taneq}) associated with the phase matching conditions of type~II SPDC is obtained from Eq~.(\ref{quartic}),
\[
y^4 -2 y^2\left[ \frac{ \gamma + 2 n_p^2(\omega; \sigma) \cos^2 \theta_i}{n_i^2 (\omega_i; \delta_i)} \right] + \frac{\gamma^2}{n_i^4(\omega_i; \delta_i)}=0,
\]
after multiplying by $(1 + \tan^2 \theta_i)^2$ and using Eq.~(\ref{clave}). For $\chi = \tan \theta_i$, the straightforward calculation yields 
\[
a_4 \chi^4 + a_3 \chi^3 + a_2 \chi^2 + a_1 \chi +a_0 =0, 
\]
which is quoted as Equation~(\ref{taneq}) in the main text, with
\bea
\begin{split}
a_0 =  -2 \xi^2  \left\{ \gamma \Delta \cos^2 \sigma 
+2   \left[ \Delta \cos ^{2} \sigma +  \dfrac{1}{n_{E}^{2}\left(\omega_i \right)} \right] n_p^2 (\omega_p; \sigma)
\right\} 
\\ + \left[ \Delta \cos^2 \sigma + \frac{2 }{n_{E}^{2}\left( \omega_i\right)} \right] \gamma^2 \Delta \cos^2 \sigma + \tau,  
\end{split}
\label{S49}
\\[1ex]
a_1 =   \left[ -  \xi^2 \left[1+2 \gamma^{-1}  n_{p}^{2}\left(\omega_p ,\sigma \right) \right]
+
\Delta \gamma \cos^2 \sigma   +   \frac{\gamma }{n_E^2 \left( \omega_i\right) } 
\right] 2 \gamma \Delta \cos \phi_i \sin (2 \sigma),
\label{S48}
\\[1ex]
\begin{split}
a_2 = - 2 \xi^2\left[  \left\{ 
\left[ \gamma  +  2  n_{p}^{2}\left(\omega_p,\sigma \right)  \right] \cos^2 \phi_i \sin^2 \sigma 
 +    \gamma\cos^{2} \sigma   \right\} \Delta + 2 \frac{n_p^2 (\omega_p; \sigma)}{n_E^2(\omega_i)} \right]
\\ + 2 \left[  \left\{ 3\Delta \cos^2 \sigma    +   \frac{ 1}{n_{E}^{2}\left( \omega_i\right)}    \right\} \cos^2 \phi_i \sin^2 \sigma
+   \frac{ \cos^2 \sigma}{n_{E}^{2}\left(\omega_i\right)} 
\right] \gamma^2 \Delta +   2 \tau,
\end{split}
\label{S47}
\\[1ex]
a_3 = \left[ - \xi^2  + \Delta \gamma \cos^2 \phi_i \sin^2\sigma +  \frac{\gamma}{n_E^2 \left( \omega_i\right) } \right]
2 \gamma \Delta \cos \phi_i \sin (2\sigma),
\label{S46}
\\[1ex]
a_4 = \left[ -  2 \xi^2 \gamma \Delta  + \left[ \Delta \cos^2 \phi_i \sin^2 \sigma   +   \frac{2}{n_{E}^{2}\left( \omega_i\right) }  \right] \gamma^2 \Delta \right] \cos^2 \phi_i \sin^2 \sigma + \tau,
\label{S45}
\eea
and
\be
\tau = \left[ \xi^2 - \frac{\gamma}{n_E^2 (\omega_i)} \right]^2.
\ee
The five coefficients $a_k$, $k=0,1,2,3,4$, as well as $\tau$, are determined in terms of the azimuthal angle $\phi_i$, the orientation $\sigma$ of the optic axis, and the three angular frequencies $\omega_p$, $\omega_s$, and $\omega_i$.

\section{General solution}

It is suitable to rewrite (\ref{taneq}) in its monic form (\ref{monic}),
\[
\chi^{4}+b_{3} \chi^{3}+b_{2} \chi^{2}+b_{1}\chi+b_{0} = 0, \quad b_k = \frac{a_k}{a_4}, \quad k=0,1,2,3,
\]
in order to make the substitution
\be
\chi = z -\tfrac14 b_3
\label{chz}
\ee 
and arrive at the standard form
\be
z^4 + p z^2 + q z + r = 0,
\label{standard}
\ee
with
\be
p =b_{2} -\tfrac38 b_3^2, \quad q =\tfrac18 b_3^3 -\tfrac12 b_{3} b_{2} +b_{1}, \quad r = - \tfrac{3}{256}b_{3}^{4} + \tfrac{1}{16} b_{3}^{2} b_{2} -\tfrac{1}{4}b_{3} b_{1} +b_{0}.
\ee

In general, for $q \neq 0$ and $r \neq 0$, we propose the change
\be
z = \alpha v,
\label{change}
\ee
so that the structure of (\ref{standard}) is preserved but the coefficients are changed 
\be
v^4 + \left( \frac{p}{\alpha^2} \right) v^2 + \left( \frac{q}{\alpha^3} \right) v + \frac{r}{\alpha^4}=0.
\label{standard2}
\ee
Formally, (\ref{standard}) and (\ref{standard2}) are equivalent, so they admit exactly the same method of solution. What is gained with the change (\ref{change}) is the parametrization of the coefficients in terms of $\alpha$, which is to be determined.

We will make the identification of both quartic equations, (\ref{standard}) and (\ref{standard2}), with the relationship
\be
(\zeta^2 +A)^2  = (\zeta +B)^2,
\label{pase}
\ee
which is solvable through the quadratic forms
\be
\zeta^2 \pm \zeta + A \pm B =0.
\ee
That is, we have at hand the roots
\be
\zeta_{1,2} = -\tfrac12  \pm \tfrac12 \sqrt{1 -4 (A+B)}, \quad 
\zeta_{3,4} = \tfrac12  \pm \tfrac12 \sqrt{1 -4 (A-B)}.
\label{zetas}
\ee
To set $A$ and $B$, let us develop the squares of (\ref{pase}), 
\be
\zeta^4 + (2A-1) \zeta^2 - 2 B \zeta + A^2 -B^2 =0.
\label{pase2}
\ee
Comparing (\ref{pase2}) with  (\ref{standard}) and (\ref{standard2}) gives respectively
\be
A = \tfrac12 (p+1), \quad B =-\tfrac12 q, \quad A^2 -B^2 = r,
\ee
and
\be
\widetilde A = \frac{p+ \alpha^2 }{2\alpha^2}, \quad \widetilde B =-\frac{q}{2 \alpha^3}, \quad \widetilde A^2 - \widetilde B^2 = \frac{r}{\alpha^4}.
\label{change2}
\ee
Then, with $\zeta = v$ in (\ref{zetas}) we obtain
\be
v_{1,2} = -\tfrac12  \pm \tfrac12 \sqrt{1 -4 (\widetilde A+ \widetilde B)}, \quad
v_{3,4} = \tfrac12  \pm \tfrac12 \sqrt{1 -4 (\widetilde A- \widetilde B)}.
\label{uves}
\ee
The introduction of (\ref{uves}) into (\ref{change}) provides the roots we are looking for
\be
z_{1,2} = -\tfrac{\alpha}{2}  \pm \tfrac12 \left[ -2p -\alpha^2 + 2q \alpha^{-1} \right]^{1/2}, \quad 
z_{3,4} = \tfrac{\alpha}{2}  \pm \tfrac12 \left[ -2p -\alpha^2 - 2q \alpha^{-1} \right]^{1/2}.
\label{zetas2}
\ee
The roots (\ref{set}) included in the main text are obtained after introducing (\ref{zetas2}) in (\ref{chz}), with $\alpha^2 = \eta$.

The parameter $\alpha$ is determined from the quadratic difference of $\widetilde A$ and $\widetilde B$ in Eq.~(\ref{change2}). Indeed, the straightforward calculation yields 
\be
\alpha^6 +2p \alpha^4 + ( p^2 -4r ) \alpha^2 - q^2 =0,
\ee
which coincides with the cubic equation (\ref{para3}) of the main text after the change $\alpha^2 = \eta$.

\section{Reduced cases}

Two special configurations are easily identified as reduced cases of the previous results.

$\bullet$ {\bf Biquadratic equation.} If $q=0$, the equation
\be
z^4 + p z^2 + r = 0,
\ee
is solved after the change $u= z^2$, and providing the roots of the quadratic equation
\be
u^2 + p u + r = 0.
\ee
That is, from the quadratic formula
\be
u_{\pm} = \tfrac12 \left[-p \pm \sqrt{p^2 -4r} \right],
\label{noventa1}
\ee
we have
\be
\chi_1 = \sqrt{u_+} - \tfrac14 b_3, \quad \chi_2 = - \sqrt{u_+} - \tfrac14 b_3, \quad \chi_3 = \sqrt{u_-} - \tfrac14 b_3, \quad \chi_4 =- \sqrt{u_-} - \tfrac14 b_3,
\label{noventa2}
\ee
where (\ref{chz}) has been used.

$\bullet$ {\bf Cubic equation.} If $r=0$ then $z=0$ is a firs root and (\ref{standard}) is reduced to the cubic equation
\be
z^3 + p z + q  = 0.
\label{cubic}
\ee
The solutions $\chi_k$ are obtained from the roots of (\ref{cubic}), together with $z=0$, after using (\ref{chz}).

\section*{Acknowledgment}

This research has been funded by Consejo Nacional de Ciencia y Tecnolog\'ia (CONACyT), Mexico, Grant Numbers A1-S-24569 and CF19-304307.


\end{document}